\documentclass[aps,floatfix,eqsecnum,showpacs,preprint,tightenlines,nofootinbib]{revtex4-1}
\usepackage{epsfig}

\usepackage{amsmath}
\usepackage{color}

\begin{document}



\title{Nucleon-deuteron scattering with the JISP16 potential}
\author{R. Skibi{\'n}ski}
\affiliation{M. Smoluchowski Institute of Physics, Jagiellonian University, PL-30348 Krak\'ow, Poland}
\author{J. Golak}
\affiliation{M. Smoluchowski Institute of Physics, Jagiellonian University, PL-30348 Krak\'ow, Poland}
\author{K. Topolnicki}
\affiliation{M. Smoluchowski Institute of Physics, Jagiellonian University, PL-30348 Krak\'ow, Poland}
\author{H. Wita{\l}a}
\affiliation{M. Smoluchowski Institute of Physics, Jagiellonian University, PL-30348 Krak\'ow, Poland}
\author{Yu. Volkotrub}
\affiliation{M. Smoluchowski Institute of Physics, Jagiellonian University, PL-30348 Krak\'ow, Poland}
\author{H. Kamada}
\affiliation{Department of Physics, Faculty of Engineering, Kyushu Institute of Technology, Kitakyushu 804-8550, Japan}
\author{A. M. Shirokov}
\affiliation{Skobeltsyn Institute of Nuclear Physics, Moscow State University, Moscow 119991, Russia}
\affiliation{Department of Physics, Pacific National University, Khabarovsk 680035, Russia}
\author{R. Okamoto}
\affiliation{Senior Academy,
Kyushu Institute of Technology, Kitakyushu 804-8550, Japan} 
\author{K. Suzuki}
\affiliation{Senior Academy,
Kyushu Institute of Technology, Kitakyushu 804-8550, Japan} 
\author{J. P. Vary}
\affiliation{Department of Physics and Astronomy, Iowa State University, Ames, IA 50011-3160, USA}

\date{\today}

\begin{abstract}

The nucleon-nucleon J-matrix Inverse Scattering Potential JISP16 is applied 
to elastic nucleon-deuteron scattering and the deuteron breakup process 
at the laboratory nucleon energies up 
to 135 MeV.
The formalism of the Faddeev equations is used to obtain three-nucleon 
scattering states. We compare predictions based on the JISP16 force 
with data and with results based on various two-body interactions, including 
the CD Bonn, the Argonne AV18, 
the chiral force with the semi-local regularization at the fifth order of the 
chiral expansion and with low-momentum interactions
obtained from the CD Bonn force
as well as with the predictions from the combination of the 
AV18 NN interaction and the Urbana IX 3N force. 
JISP16 provides a satisfactory description of some observables 
at low energies but strong deviations from data 
as well as from standard and chiral potential predictions 
with increasing energy.
 However, there are also polarization observables at low energies for which 
the JISP16 predictions differ from 
those based on the other forces 
by a factor of two. The reason for such a behavior can be traced back 
 to the P-wave components of the JISP16 force. At higher energies 
the deviations can be 
enhanced by an 
interference with higher partial waves 
and by the properties of the JISP16 deuteron wave function. 
In addition, we compare the energy and angular dependence 
of predictions based on the JISP16 force 
with the
 results of  the low-momentum 
interactions obtained with different values of the momentum cutoff parameter. 
We found that such low-momentum forces 
can be employed to interpret the nucleon-deuteron elastic scattering data only 
below some specific energy which depends on the cutoff parameter.  
Since JISP16 is defined in a finite oscillator basis, it has properties similar to
 low momentum interactions and its application 
to 
the description of nucleon-deuteron
scattering data is
limited to a low momentum transfer 
region. 
 
\end{abstract}

\pacs{21.45.-v, 13.75.Cs, 25.40.Cm}

\maketitle


\section{Introduction}

Various models of the nucleon-nucleon (NN) interaction have been derived in the past.
Some of them, like the Charge Dependent Bonn (CD Bonn)~\cite{CDBonn, Mac01}  potential 
arise from the boson-exchange picture of nuclear interactions and aim not only at describing data 
but also at providing insight into the underlying physics. 
Other models, like the Argonne AV18~\cite{AV18} potential,
exploit the possible operator structure and
 preserve the pion exchange  picture for the long-range component of the interaction
while introducing 
a 
phenomenological parametrization for the short-range part. 
Even the most advanced models 
have to incorporate 
numerous adjustable parameters (about 40 for the CD Bonn or the AV18 models) 
whose values have to be fixed from NN data. 
Such semi-phenomenological potentials describe the NN 
scattering data and deuteron properties with high precision, 
achieving for example $\chi^2/{\rm data'99}$=1.35 in the case of the AV18 force  
and $\chi^2/{\rm data'99}$=1.01 for the CD Bonn model~\cite{Machleidt2001}.

Another 
approach arises 
from the chiral effective field theory 
where one builds the effective nuclear potential basing on Lagrangian 
for nucleon and pion fields ~\cite{Machleidt_review, Epelbaum_review}. So far
the chiral NN interaction has been derived completely up to the fifth order
of the chiral expansion (N$^4$LO)~\cite{imp1,imp2}.
Moreover, the dominant contributions at the sixth order are 
also known~\cite{Entem_6th}.
The good quality of the chiral force with the semi-local 
regularization~\cite{imp1,imp2} and weak cutoff dependence of predictions 
 was confirmed in the
3N continuum, both for strong~\cite{lenpic3} 
and electroweak~\cite{Skibinski_electroweak} processes,
as well as
in nuclear structure calculations. 

The JISP16 NN potential was presented ten years ago in 
Ref.~\cite{Shirokov_JISP16}. 
This force was a successor of the J-matrix Inverse Scattering 
Potential JISP6~\cite{Shirokov_JISP6}, which in turn followed
 the Inverse Scattering 
Tridiagonal Potential (ISTP) developed within 
the J-matrix
inverse scattering 
formalism in~\cite{Shirokov_ISTP}. 
Free parameters of the JISP6 
force have been fixed by fitting to the NN phase shifts as well as to  bound 
and resonance states of nuclei up to $A=6$.
Correspondingly, bound and resonance states of nuclei up to $^{16}$O have been
 used to adjust the free parameters
of its successor, the JISP16 force.
Both JISP forces describe also NN scattering data with high precision, 
comparable to 
the other modern potentials, reaching $\chi^2=1.03(1.05)$ with 
the neutron-proton data'1992(1999)~\cite{Stoks}(~\cite{Mac01}).
The JISP forces assume charge independence
and, regarding the NN system, 
only the neutron-proton scattering data and the deuteron properties
have been taken into account 
when fixing the parameters.

An  
important feature of the JISP16 force is that it provides sufficient 
convergence of the No-Core Shell Model~\cite{BarrettNCSM} 
calculations 
enabling 
accurate predictions for nuclear binding 
energies and spectra of excited nuclear states  
with established 
extrapolation techniques~\cite{VMS, Coon,Sid16} and 
the ability to perform perturbative calculations of nuclear 
matter~\cite{applicationJISPinNM}. The description of properties of light nuclei by JISP16
is rather accurate, see Refs.~\cite{Shirokov_review,IJMPE-2013}. 
In particular, the accuracy of $^{14}$F binding energy and spectrum  predictions ~\cite{14F} 
based on this interaction
was later confirmed by the first experimental study of this nucleus in Ref.~\cite{14Fexp}.
The JISP16 force is also known to provide an accurate {\em ab initio} description of resonance energies 
and widths in nucleon-$\alpha$ scattering~\cite{PRC79,ApplMathInfSci,n-alpha} and in the
tetraneutron system~\cite{4n}.
These successful applications of the JISP16 interaction 
have encouraged 
us to test this force also in the studies of nucleon-deuteron scattering which is a well-known
challenge for inter-nucleon forces~\cite{N-d-review,lenpic1,lenpic_Golak}. 

Momentum space matrix elements of the JISP16 potential decrease quickly 
with increasing momenta which 
makes this force very useful 
in nuclear structure calculations. This 
welcome 
feature of NN interactions 
was 
one of the reasons for developing the so-called
``low momentum interactions" 
$V_{\rm{low}\;{\rm k}}$~\cite{Bogner1, Bogner2, Fujii, Nogga}. 
The force of Refs.~\cite{Bogner1, Bogner2}, which originated 
from an application of 
the regularization group methods to soften the interaction, 
has also been widely used in calculations of energy levels of 
various nuclei and in nuclear matter studies (see reviews~\cite{Vlowk-rev,Kuo-review} 
and references therein).
These methods take care about the unitarity of the transformation both 
in NN and many-nucleon systems. They thus preserve the description of two-nucleon 
(2N)\footnote{We use NN and 2N interchangeably with preference for the latter when warranted by the context.} 
and 3N observables both in bound states and scattering processes with arbitrary energies. 
Another idea lies at the heart of the low momentum interaction obtained in Ref.~\cite{Fujii}. 
There the NN force was constructed by means of a transformation which 
cuts off the short range (or equivalently the high momentum) part of the realistic 
input potential. As a consequence it retains the same description of the NN observables 
but only for the c.m. NN energy below the value defined by the cut off parameter.

The authors of Refs.~\cite{Bogner1, Bogner2} recommend using the 
cutoff parameter around $\Lambda$=2~fm$^{-1}$,
in contrast with Ref.~\cite{Fujii}, where  the value of $\Lambda$=5~fm$^{-1}$ was
suggested. 
 It should be emphasized that in Refs.~\cite{Bogner1, Bogner2,Fujii} 
additional 3N forces, which appear when on-the-energy-shell equivalent 
low momentum 2N
 interactions are used 
 in systems with more than two particles \cite{jurgenson2009,furnstahl2013}, 
 were omitted. 
 It is clear that the value of $\Lambda$ correlates with 
the energy range where the corresponding 
$V_{\rm{low}\;{\rm k}}$ can be used. 
In the few-nucleon sector, the $V_{\rm{low}\;{\rm k}}$ interaction constructed 
within the  
approach of Refs.~\cite{Bogner1, Bogner2}
from various models of the NN interaction, was applied 
in Ref.~\cite{Deltuva} to study the neutron analyzing power in elastic 
neutron-deuteron scattering at the neutron energy of 3~MeV,
the breakup cross section in the space star (SST) configuration at $E=13$~MeV  
and some selected observables in neutron-triton scattering
at energies below 6 MeV. 
The application of the $V_{\rm{low}\;{\rm k}}$ potential obtained using 
the method of Ref.~\cite{Bogner2} to 
proton-deuteron elastic scattering was presented also 
in Ref.~\cite{Marcucci2009} at low center-of-mass energies, up to 2 MeV.
At these energies the used $V_{\rm{low}\;{\rm k}}$ force (based on the AV18 with 
the cutoff parameters $\Lambda$ equal to 2.2~fm$^{-1}$)
delivers a very good description of the cross section and various 
spin observables. 

To the best of our knowledge, up to now the $V_{\rm{low}\;{\rm k}}$ potential 
obtained using the approach of Ref.~\cite{Fujii} 
has not been used to study elastic nucleon-deuteron (Nd) scattering. 
We fill this gap in the present paper and
show predictions for various observables in Nd scattering at the laboratory 
kinetic energies of the incoming 
nucleon ranging from 5~MeV to 135~MeV. Our results are obtained with the $V_{\rm{low}\;{\rm k}}$ 
force 
derived with the method of Ref.~\cite{Fujii} applied to the CD Bonn force. 
We use the $\Lambda$ cutoff values
ranging from 1.5~fm$^{-1}$ to 5.0~fm$^{-1}$. Observables obtained in this 
way are compared 
with the CD Bonn predictions and
with the JISP16 results.
Being aware that the additional three-nucleon force should be 
taken into account when applying the low momentum interaction of Ref.~\cite{Fujii}
to 3N processes, we use here this interaction in order to compare it with 
the JISP16 force rather than to describe specific 3N data. 

The role of the induced 3NF resulting from the regularization 
group methods has been investigated both in the nuclear structure, 
see e.g. Ref.\cite{jurgenson2009}, and at low energies in 3N scattering~\cite{Deltuva}, while
the role of additional 3NF accompanying the low momentum interaction of Ref.~\cite{Fujii}
has been estimated only for the $^3$H and $^4$He nuclei in Ref.~\cite{Fujii}.
It has been found that one can expect the contribution of additional 3NF to 
the triton binding energy up to 0.7 MeV, depending on the cut-off parameter value 
in the range above 1.0 fm$^{-1}$. This is approximately 50\% of
the contribution given by "realistic" 3NFs. Thus it is plausible to think
that also in the Nd scattering process the impact of additional 3NF 
is smaller than the contribution of "realistic" 3NFs. 

The JISP16
potential was, from the very beginning, assumed to give a good description 
of nuclei in absence
of the  
many-body interactions and 
many-body observables
have been used to fix its parameters. 
Indeed, results from the structure calculations confirm this feature
of the JISP16 potential. In this paper we would like to check if 
this observation is 
also valid for the 3N 
scattering observables. To this end 
we compare the predictions 
based on the JISP16 force with ones obtained from a Hamiltonian 
containing an explicit 3NF,
namely by 
using the AV18 NN potential combined with the Urbana IX~\cite{UrbanaIX} 3NF. 
However, we are also interested in a comparison of predictions based on 
the JISP16 force 
with the results based on other models of the 2N interaction 
with a focus on the role of softening those interactions.
   
The paper is organized as follows: in the next section we shortly describe 
the framework of the 3N
Faddeev equations.
In Sec.~\ref{Sec_bound} we discuss some properties of 2N and 3N bound states.
Predictions for the Nd
elastic scattering and the deuteron breakup reaction obtained, 
for the first time, with the JISP16 potential,
 and their 
comparison with results of
calculations based on semi-phenomenological and chiral N$^4$LO potentials 
are shown and discussed 
in Sec.~\ref{Sec_Ndscattering}. 
In Sec.~\ref{Vlowk} we compare the JISP16 Nd scattering results with those 
based on the $V_{\rm{low}\;{\rm k}}$ forces.
Finally, we present our summary and conclusions 
in Sec.~\ref{Summary}.

\section{Formalism}

Results for the Nd 
scattering 
 presented in this work have been obtained in the framework of the 
Faddeev equations in momentum space. Since this formalism is nowadays one of the standard 
techniques to investigate 3N reactions and has been described in detail many times, 
we only briefly remind the reader the most fundamental steps. The interested reader can find more 
details, e.\,g., in Refs.~\cite{Glockle_book, Glockle_raport, Witala}.

In this approach, the Faddeev equation for an auxiliary state $T \vert \phi \rangle $ is 
the central 
equation to be solved. It reads
\begin{equation}
T \vert \phi \rangle = tP \vert \phi \rangle + tPG_0 T \vert \phi \rangle + (1 + t G_0 )V_4^{(1)} (1 + P)\vert \phi \rangle
+ (1 + tG_0 )V_4^{(1)} (1 + P)T \vert \phi \rangle,
\label{eq_Fadd_NN_3N}
\end{equation}
where the initial state $\vert \phi \rangle$ is composed of a deuteron and a relative momentum eigenstate of the projectile nucleon,
$P$ is a permutation operator which takes into account the
identity of the nucleons and $G_0$ is the free 3N propagator.
The 2N interaction $V$ together with the 2N free propagator $\tilde{G_0}$ 
appear in the Lippmann--Schwinger equation for the 2N $t$-matrix
\begin{equation}
t=V + V \tilde{G_0} t\;.
\label{eq_LS}
\end{equation}
In Eq.~(\ref{eq_Fadd_NN_3N}) $V_4^{(1)}$ is that part of the 3NF which is symmetrical under the exchange
of nucleons 2 and 3. 
When the 3NF is neglected, Eq.~(\ref{eq_Fadd_NN_3N}) reduces to
\begin{equation}
T \vert \phi \rangle = tP \vert \phi \rangle + tPG_0 T \vert \phi \rangle \;.
\label{eq_Fadd_NN}
\end{equation}

We solve Eqs.~(\ref{eq_Fadd_NN_3N}) and~(\ref{eq_Fadd_NN}) in the momentum space partial wave scheme.
We work with the $\vert p,q,\alpha \rangle$ states with $p=\vert \, \vec{p} \, \vert$ and $q=\vert \, \vec{q} \, \vert$
being the magnitudes of the relative Jacobi momenta $\vec{p}$ and $\vec{q}$. Further, $\alpha$ represents the set of discrete
quantum numbers for the 3N system in the $jI$-coupling
\begin{equation}
\alpha = \big( (l,s)j; (\lambda, \frac12)I; (j,I)J M_J; (t \frac12)T M_T \big)\;.
\end{equation}
Here $l$, $s$, $j$ and $t$ denote the orbital angular momentum, 
total spin, total angular momentum and total isospin of the 2-3 subsystem.
Further, $\lambda$ and $I$ are the orbital and total angular momenta of the 
spectator nucleon 1
with respect to the center of mass of the 2-3 subsystem. Finally, 
$J$, $M_J$, $T$ and $M_T$ are the total angular momentum of the 3N system,
its projection on the quantization axis, the total 3N isospin 
and its projection, respectively.

It is worth noting that during solving  Eqs.~(\ref{eq_Fadd_NN_3N}) 
or (\ref{eq_Fadd_NN}) the 2N force matrix elements, present in the $t$-operator,
clearly interfere
which can significantly affect the observables.
We solve Eq.~(\ref{eq_Fadd_NN}) by generating its Neumann series 
and summing it up by using
the Pad\'e method~\cite{Glockle_raport}.
For results presented here we use all partial waves with 
$j \leq 4$ and $J \leq \frac{25}{2}$. 
 More details about our numerical performance
can be found in~\cite{Glockle_raport}.

The JISP16 potential is initially available~\cite{JISP_web_page} 
in the harmonic oscillator (HO) 
basis
which is used commonly in nuclear structure calculations. 
The matrix elements of that potential 
in 2N momentum space basis
$\vert p, \tilde{\alpha} \rangle \equiv \vert p, (l,s) j 
; t  
\rangle$ 
are then given by 

\begin{equation}
\langle p',\tilde{\alpha}' \vert V \vert p, \tilde{\alpha} \rangle = 
\sum_{n=0}^{n^{l}_{\max}}\; \sum_{n'=0}^{n^{l'}_{\max}}
(-1)^{n+n'} i^{l'-l} R_{n'l'}(p'b)R_{nl}(pb) b^3 \langle n',l' \vert V_{sj} \vert n,l \rangle \,, 
\end{equation}   
where 
$n$ and $n'$ are  the principal quantum numbers for HO states 
and, due to the definition of the JISP16 interaction,
$n^{l}_{\max}=(8-l)/2$ or $(9-l)/2$ depending on the parity.
Using the notation given in Appendix 1 of Ref.~\cite{Liebig}, 
the HO radial functions $R_{nl}(\rho)$ are
given by 
\begin{equation}
R_{nl}(\rho)=(-1)^n \Big[ \frac{2n!}{\Gamma(n+l+3/2)} \Big]^{\frac12} \exp(\frac{-\rho^2}{2}) \rho^l L_n^{l +\frac12}(\rho^2)\;,
\end{equation}
where $L_n^{l +\frac12}(x)$ are the generalized Laguerre polynomials, 
$\Gamma(z)$ is the Euler gamma function,
the HO length 
$b=\sqrt{\frac{\hbar^2}{m_r \hbar\Omega}}$ with $m_r=\frac12 m$ being the reduced 
mass of the 2N system, the average nucleon mass $m=\frac{m_n+m_p}{2}$ and $\hbar \Omega=40$~MeV.

Since momentum space matrix elements of the JISP16 potential are restricted to 
low momenta, we will compare its predictions to results obtained with a number 
of $V_{\rm{low}\;{\rm k}}$ potentials, whose nonzero matrix elements 
 are restricted to momenta inside intervals of decreasing size, 
 generated from the neutron-proton 
version of the high-precision CD~Bonn interaction. 
To obtain the matrix elements of the $V_{\rm low~k}$ potential, we use the \=Okubo theory~\cite{Okubo}
of the unitary transformation which splits the Hilbert space into low and high momenta subspaces.
Namely, in order to separate the momentum space to a low-momentum region and a high-momentum one, we introduce 
the following
projection operators ($P$ and $Q$)
\begin{eqnarray}
&&P=\int_0^\Lambda | p \rangle \langle p | dp,     \cr
&&Q=\int_\Lambda^\infty  | p \rangle \langle p | dp,
\end{eqnarray}
where $\Lambda$ is a momentum cutoff whose value will be specified later. 
Given the unitary transformation operator 
of the $\bar {\rm O}$kubo theory, the effective Hamiltonian $ PH'P$ in the $P$ space takes the 
form
\begin{eqnarray}
PH'P=P(1+\omega^\dagger \omega )^{-{1\over2}} (H+\omega ^\dagger H + H\omega +\omega^\dagger H \omega ) (1+\omega^\dagger \omega)^{-{1\over2}} P,
\end{eqnarray}
where the original Hamiltonian $H$ consists of the kinetic energy $H_0$ and the original potential~$V$
and where $\omega$ is a wave operator which satisfies the condition $\omega = Q\omega P$.
Then the low-momentum potential $V_{\rm low~k}$  is obtained as
\begin{eqnarray}
V_{\rm low~k}= PH'P -PH_0P.
\end{eqnarray}
The details of the two methods to obtain the wave operator $\omega$ are given in Refs.~\cite{Fujii,Suzuki,Kuo}, and in this paper we 
make use of the Suzuki Method~II~\cite{Suzuki}. 
As will be discussed in Sec.\ref{Vlowk}, the $\Lambda$ cut-off value 
determines the range of incoming nucleon energy where such a $V_{\rm low~k}$ potential is applicable. 
Even an extension of this method to the unitary transformation in 
the three-particle space and taking into account additional 3N interactions 
emerging from such a transformation, will not result in a proper description of 3N 
observables at relative initial momenta above the value given by the cut off parameter. 

In the case of the chiral interaction, we use the N$^4$LO neutron-proton force \cite{imp1,imp2} with 
the semi-local regularization induced in coordinate space by the
regulator function $f(r)  =  [1-\exp{(\,\rm{-}\,(\frac{r}{R})^2)}]^6$,
with the regulator value $R$=0.9~fm.

\section{The bound states}
\label{Sec_bound}

The neutron-proton phase shifts obtained with the JISP16 force agree very well with values extracted from experimental data
by the Nijmegen group~\cite{phase_shifts} up to the nucleon laboratory energy 350 MeV. 
Below 200 MeV the only deviation, around 15\%, is observed for the $^3S_1-{^3D_1}$ mixing 
parameter $\epsilon_1$ at energies lower than $\approx 40$~MeV.

\begin{table}
\begin{tabular}{|c|c|c|c|c|c|}
\hline
   & $E_{\rm deu}$~[MeV] & P($^3S_1$) & P($^3D_1$) & $\langle E_{\rm pot} \rangle$~[MeV] & $\langle E_{\rm kin} \rangle$~[MeV]  \\
\hline
JISP16             & $-2.2246 $ & 96.02  & 3.98  & $-12.987$  & 10.763 \\
N$^4$LO (R=0.9 fm) & $-2.2233 $ & 95.71  & 4.29  & $-21.115$  & 18.892 \\
AV18               &$ -2.2422 $ & 94.22  & 5.78  & $-22.125 $ & 18.882 \\
CD Bonn (non-rel)     & $-2.2232$  & 95.14  & 4.86  & $-17.822$  & 15.599 \\
\hline
\end{tabular}
\caption{The deuteron g.\:s. 
 energy $E_{\rm deu}$, the $^3S_1$ and $^3D_1$ state probabilities as well as 
the potential and the kinetic energy expectation values obtained with various NN interactions.}
\label{tab1}
\end{table}

\begin{figure}[b!]
\includegraphics[width=.4\textwidth,clip=true]{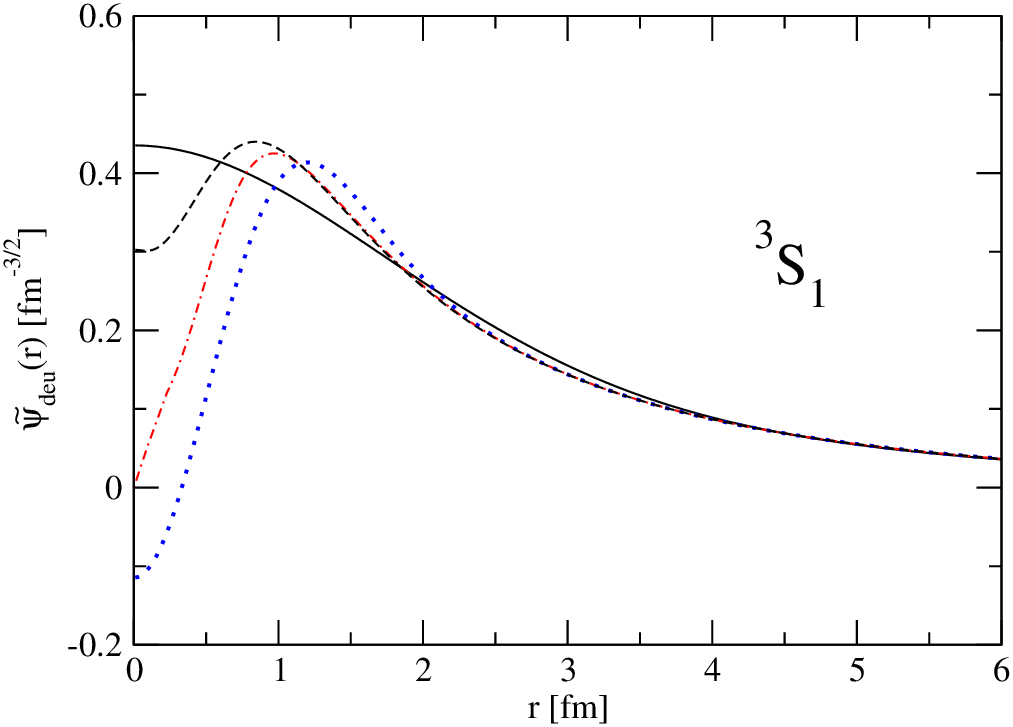}
\includegraphics[width=.4\textwidth,clip=true]{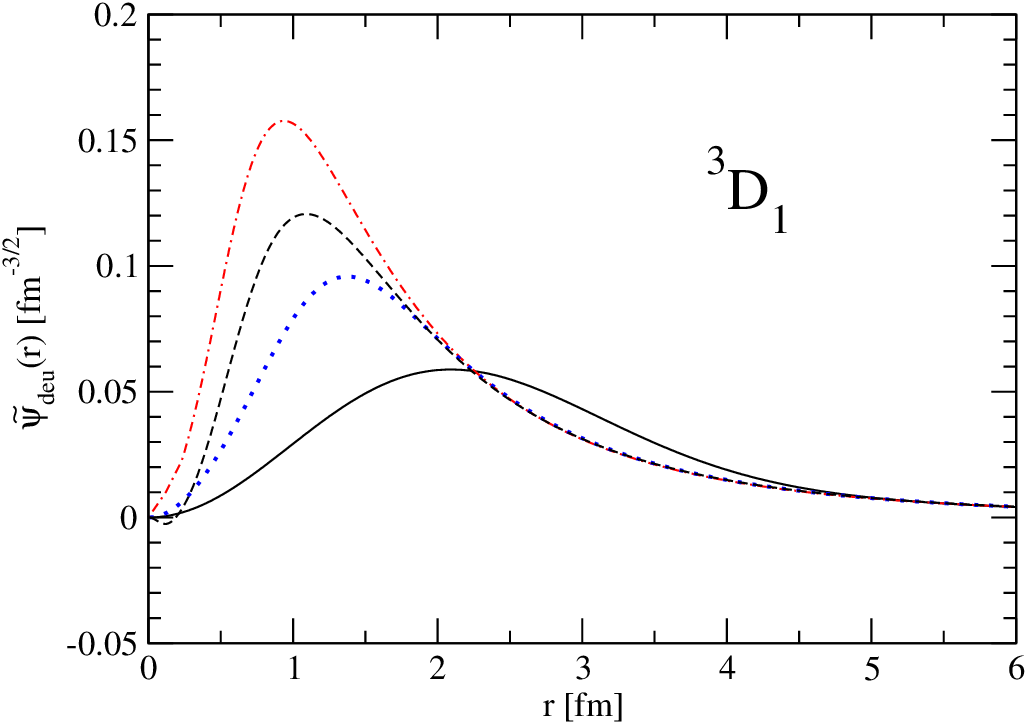}
\includegraphics[width=.4\textwidth,clip=true]{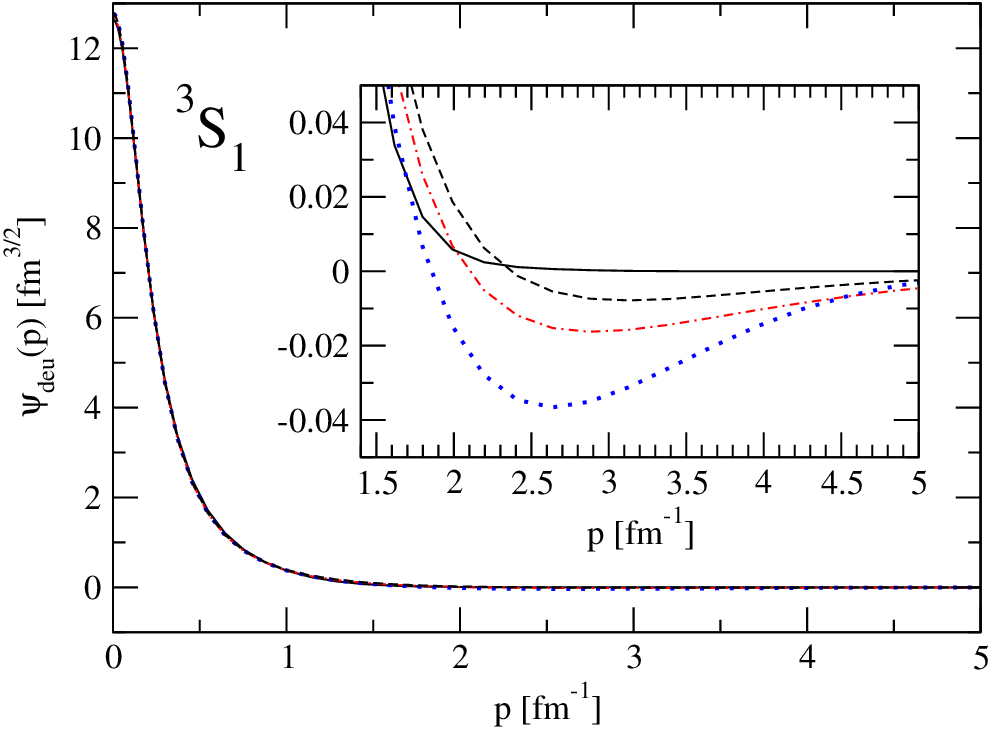}
\includegraphics[width=.4\textwidth,clip=true]{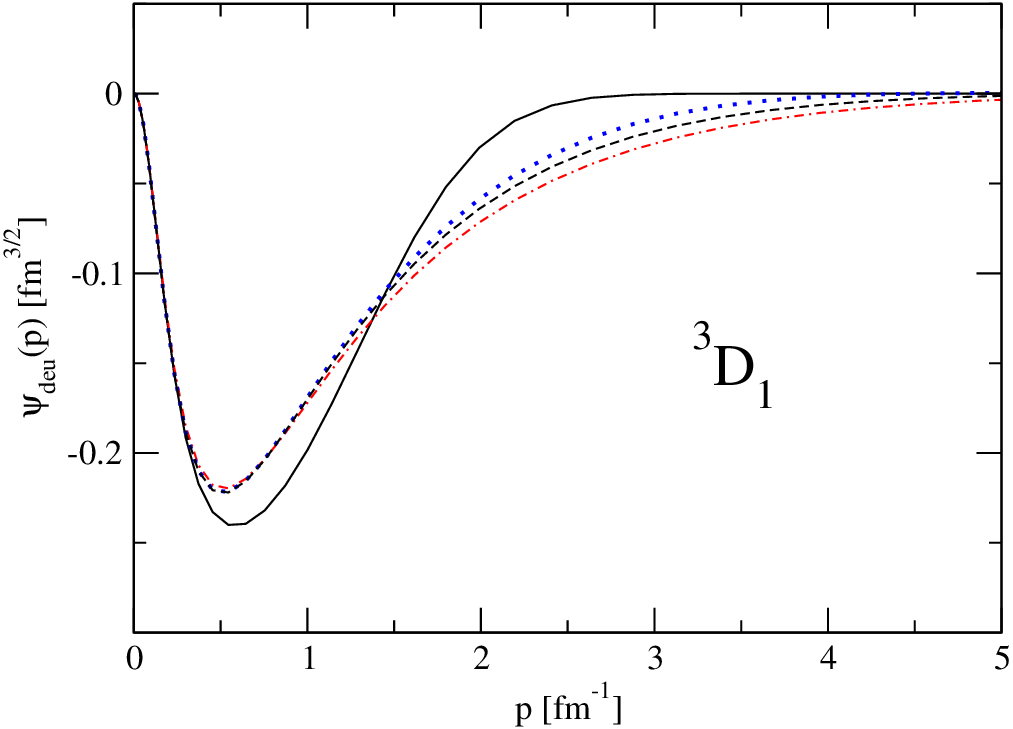}

\caption{(color online) The deuteron wave functions in coordinate $\tilde{\psi}_{deu}(r)$ (top) 
and momentum $\psi_{deu}(p)$ (bottom) space, respectively. The $^3S_1$ and the $^3D_1$ components are shown
in the left and right columns, respectively. The black solid, red dash-dotted, black dashed and blue dotted 
curves are for the JISP16, the AV18, the CD~Bonn and the chiral N$^4$LO forces, respectively.}
\label{fig1}
\end{figure}

The deuteron properties are also very well reproduced by the JISP16 
potential~\cite{SHIROKOVphasea}.
The deuteron ground state (g.\:s.) 
energy, which is one of the observables used to fix potential parameters, agrees with 
the experimental value  
$E_{\rm deu}=-2.22457$ MeV for the JISP16 as well as for  other 
 NN interaction models. 
In Tab.~\ref{tab1} we give a few additional quantities 
 which are not observables but which shed some light on
 the deuteron properties
 arising from various forces.
The  $^3D_1$ state probability takes the smallest value for the JISP16 
force but remains close to numbers obtained with the other models. 
 The  expectation values of the potential and  kinetic energy in the deuteron 
 for the JISP16 model differ significantly from the corresponding numbers 
 obtained with other interactions.
That difference can be understood after examining the deuteron wave functions 
given by these potentials, which are shown in Fig.~\ref{fig1}, 
in coordinate ($\tilde{\psi}_{deu}(r)$) and 
momentum ($\psi_{deu}(p)$) spaces. 
The $^3S_1$ component of the JISP16 deuteron wave function in coordinate space
decreases monotonically contrary to the $^3S_1$ wave functions for
 other NN interactions, which all have a maximum 
 around $r=1$~fm (see  Fig.~\ref{fig1}).
All $^3D_1$ state wave functions have a maximum, which for the JISP16 is 
localized  at $r \approx 2$~fm while for the other potentials at $r \approx 1$~fm. 
In momentum space both components of the deuteron wave function for the JISP16 
interaction are
compressed to smaller momenta compared to the other potential predictions. 
This is clearly seen also in the inset in
the left column and the bottom row of Fig.~\ref{fig1}. 
For the JISP16 force the $^3S_1$ $\psi_{deu}(p)$ drops
quickly and becomes negligible above $p \approx 2.5$~fm$^{-1}$. 
The $^3S_1$ deuteron wave function for the other potentials shows a shallow
minimum at these momenta and approaches zero at momenta above $p$=5~fm$^{-1}$. 
The differences in the behavior of  $^3S_1$ wave functions explain the 
 energy expectation values, shown in Tab.~\ref{tab1}---the wave function
localized at lower 
momenta (see lower panels of Fig.~\ref{fig1})
leads to a smaller expectation  value of
 the kinetic and thus also of the potential energy.

Predictions for the triton binding energy as well as for 
the kinetic and potential energy expectation values in the triton 
are presented in Tab.~\ref{tab2}.
The $^3$H  g.\:s. 
 energy obtained with the JISP16 interaction is much 
closer to the experimental value ($-8.482$ MeV~\cite{3H_exp})
than predictions based on the other NN forces. 
This is another example where 
the pairwise 
 JISP16 model alone works well in nuclear structure calculations
 without resorting to additional 3N dynamics. For the other potentials 
listed in   
 Tab.~\ref{tab2} one needs explicit 3NF's to explain the triton binding 
energy as exemplified  
for the AV18+Urbana IX combination. 
The expectation values of the kinetic energy and the two-body part of the 
potential energy operators in the triton show the same tendencies 
as in the deuteron case: again the values obtained by using 
the JISP16 force are significantly smaller than ones for the other models. 
Comparing the g.\:s. energies given in Tab.~\ref{tab2} 
with experimental values one has to be aware that the JISP16 model neglects
charge independence breaking of the NN force. 
However one can think about the JISP16 
force as about an effective interaction with parameters fixed by 
data from many-nucleon systems, which may include some 
effects of charge independence breaking.
The effect of mimicking the charge dependence in many-body systems 
by off-shell properties of charge-independent NN interaction is more pronounced 
in another 
interaction fitted to many-nucleon systems,  
the Daejeon16~\cite{Dae16} which reproduces the binding energies of not only $N=Z$ nuclei 
but also of nuclei with large difference of numbers of neutrons and protons 
like, e.g. $^{10}$He.

The possibility of avoiding explicit  3NF's 
 when using the two-body JISP16 
 interaction for explaining nuclear binding energies prompted us to 
check if this strategy could be successful 
in Nd 
scattering. 
It would be an interesting finding, since up to now while
using high precision, standard  or chiral NN potentials,  
 resorting to explicit 3NF's is mandatory at higher 
energies in order to explain data for some scattering observables. 
 
\begin{table}
\begin{tabular}{|c|c|c|c|c|}
\hline
   & $E_{\rm 3H}$~[MeV] & $\langle E_{\rm pot}^{(NN)} \rangle$~[MeV] & $\langle E_{\rm pot}^{(3N)} \rangle$~[MeV]  &$\langle E_{\rm kin} \rangle$~[MeV]  \\
\hline
JISP16             &$ -8.369 $ & $-35.766$ & $-$ & 27.399 \\
N$^4$LO ($R=0.9$ fm) & $-7.828$  & $-56.952$ & $- $& 49.124 \\
AV18               & $-7.656 $ & $-54.461$ & $-$ & 46.805 \\
AV18+Urbana IX      & $-8.507 $ & $-58.686$ & $-1.123$  & 51.304 \\
\hline
\end{tabular}
\caption{The $^3$H g.\:s.  
energy $E_{\rm{3H}}$ and the expectation values for the 2N potential energy ($E_{\rm pot}^{(NN)}$), 
the 3N potential energy ($E_{\rm pot}^{(3N)}$) and the kinetic energy ($E_{\rm kin}$) obtained with various NN or NN+3N interactions.} 
\label{tab2}
\end{table}

\section{ \boldmath$\rm Nd$ scattering with JISP16, semi-phe\-no\-me\-no\-lo\-gical and chiral forces}
\label{Sec_Ndscattering}

In this section we present predictions obtained with the JISP16, 
the semi-phe\-no\-me\-no\-lo\-gical and the chiral N$^4$LO NN forces 
for various 
observables in the Nd scattering process at
incoming nucleon 
laboratory energies $E=5$, 13, 65 and 135 MeV.
Since the JISP16 potential has been derived only for the neutron-proton 
system, we
use also the neutron-proton version of the corresponding NN interaction 
in calculations with other forces. 
In some of the following figures we compare our predictions also to the proton-deuteron data, 
what is justified by small effects
of the Coulomb force (neglected in the theoretical calculations). 
Such effects are only visible at lower
energies and forward scattering angles~\cite{Deltuva_pd}.

\begin{figure}
\includegraphics[width=.8\textwidth,clip=true]{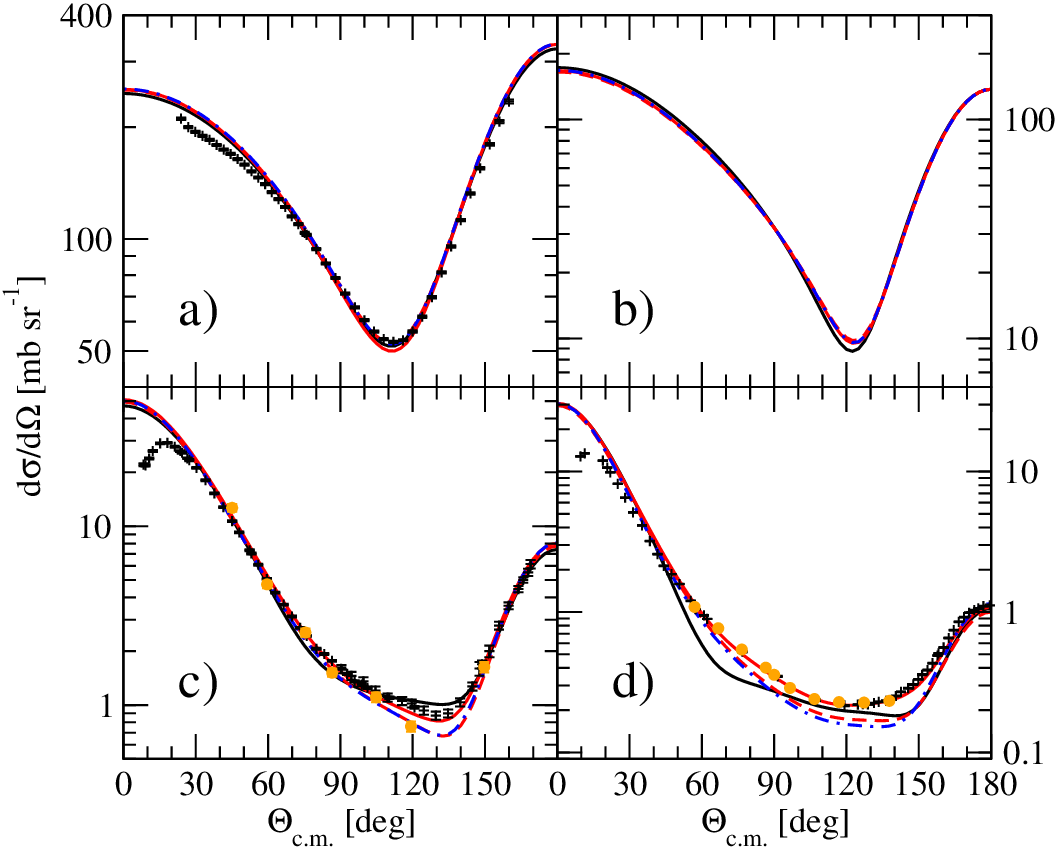}
\caption{(color online) The differential cross section d$\sigma$/d$\Omega$~[mb~sr$^{-1}$] for elastic Nd scattering at the
incoming nucleon laboratory  energy a) $E=5$ MeV, b) $E=13$ MeV, c) $E=65$ MeV and d) $E=135$~MeV as a function of the center-of-mass 
scattering angle $\theta_{\rm{c.m.}}$. The black solid, red dashed, red solid and blue dash-dotted curves represent
predictions based on the JISP16, AV18, AV18+Urbana IX and chiral N$^4$LO (with the regularization parameter $R=0.9$ fm) forces, respectively.
The data are in a) from Ref.~\cite{exp5} ($pd$ pluses), in c) from Ref.~\cite{Shimizu} ($pd$ pluses) and~\cite{Ruhl} ($nd$ orange circles) and 
in d) from Ref.~\cite{exp135_1} ($pd$ pluses) and~\cite{exp135_2} ($pd$ orange circles).}
\label{fig2} 
\end{figure}

Differential cross sections at the  above listed energies are shown in Fig.~\ref{fig2}. 
At $E=5$~MeV the predictions based on the JISP16 cross section 
(black solid curve) practically do not differ 
from those for the AV18 (the red dashed curve) or the chiral N$^4$LO 
(the blue dash-dotted curve) forces. 
They are also very close to the predictions based on the AV18 NN potential 
combined with  
the Urbana IX 3N force indicating that at this energy 3NF effects 
are practically negligible for the elastic scattering cross section. 
This picture changes at higher energies. At $E=13$ MeV in the region around the
 minimum of 
 the cross section the JISP16 predictions 
are approximately 10\% below the other practically overlaping predictions.
 Since also here the AV18 and AV18+Urbana IX results are nearly the same, 
we conclude that 3NF effects in the elastic scattering cross section
are negligible for this energy, too. At $E=65$ MeV the JISP16 predictions in the same region of angles
  are approximately 15\% above the data. 
 Here AV18+Urbana IX prediction
describes
the data well while the results of the other two NN potentials overlap and 
clearly underestimate the experimental values.
 At this and at higher energies the 3NF plays an important role in the region 
around the minimum 
of the cross section---the AV18 predictions are substantially increased by 
taking  the Urbana IX force into account.
The JISP16 predictions are, when compared to the other NN based results, 
 shifted in the same direction as AV18+Urbana~IX predictions,
however this shift is too strong and results in overestimating the data. 
At $E=135$~MeV the deviation of the JISP16 predictions from the others 
 increases significantly. While in the minimum of the cross section 
the JISP16 results are closer to the data than those for the AV18 and 
the chiral forces, at scattering angles
$\theta_{\rm{c.m.}}$ in the range between 50$^{\circ}$ and 80$^\circ$ one observes 
a serious discrepancy from the data and from the other predictions. It is 
interesting to note that at this energy  
all cross section predictions overlap up to $\theta_{c.m.} \approx 50^{\circ}$.  

\begin{figure}
\includegraphics[width=.8\textwidth,clip=true]{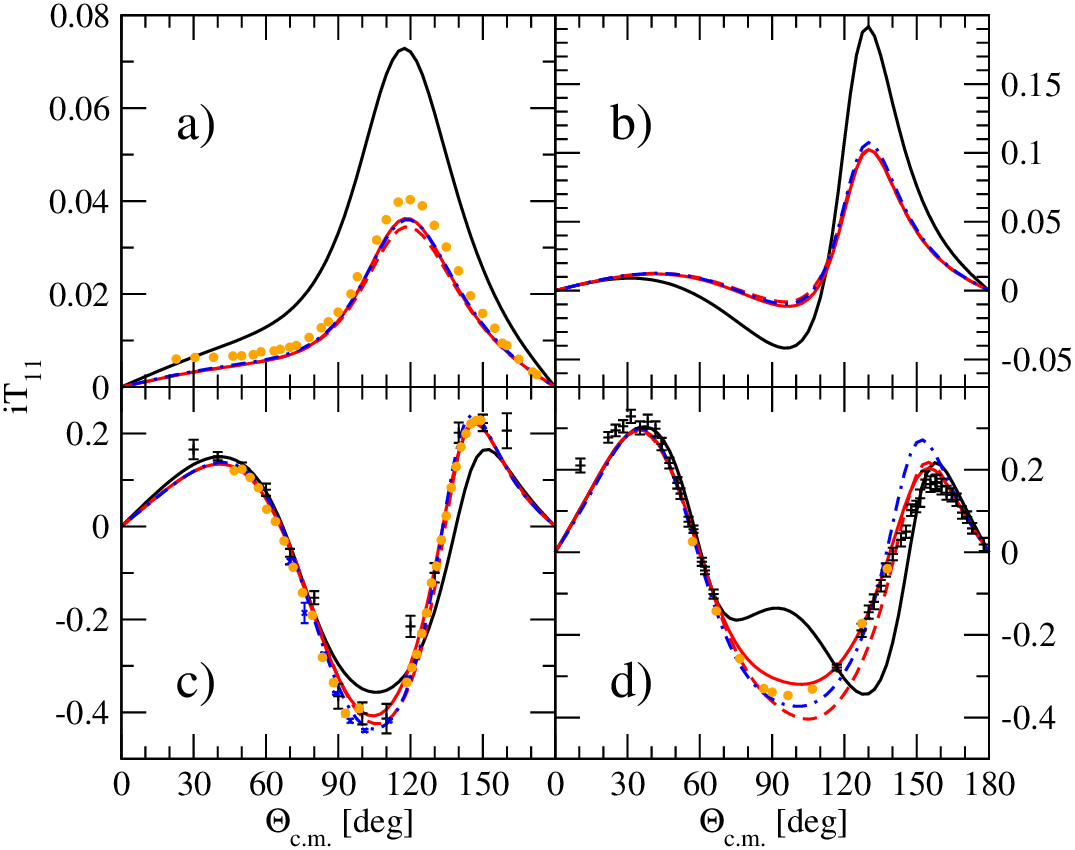}
\caption{(color online) The deuteron analyzing power iT$_{11}$ for elastic 
Nd scattering at the
incoming nucleon laboratory energy a) E=5 MeV, b) E=13 MeV, c) E=65 MeV 
and d) E=135 MeV as a function of the center-of-mass
scattering angle $\theta_{\rm{c.m.}}$. Curves are as in Fig.~\ref{fig2}.
The data in a) are from Ref.~\cite{Sowinski} ($pd$ orange circles), in c) are 
from Ref.~\cite{Witala_it11} ($pd$ pluses),
~\cite{Stephan} ($pd$ orange circles) and~\cite{Mardanpour} ($pd$ blue x-es), 
and in d) from 
Ref.~\cite{exp135_1} ($pd$ pluses) and~\cite{exp135_2} ($pd$ orange circles).
}
\label{fig4}
\end{figure}

Fig.~\ref{fig4} presents the deuteron vector analyzing power iT$_{11}$ 
as a representative for vector analyzing powers. Contrary to the cross section, 
the iT$_{11}$ is 
the observable 
for which the JISP16 model fails completely even at the lowest energy.
The predictions based on the JISP16 force at $E=5$~MeV and $E=13$ MeV, are 
about twice as large as predictions obtained with the other interactions.
Despite the big difference in the magnitudes, the shapes of the iT$_{11}$ 
 curves are similar.
At $E=65$ MeV the JISP16 predictions come closest to 
those of the other interactions, which agree well with the data, while 
diverging from them around the minimum of the analyzing 
power and around $\theta_{\rm{c.m.}}=145^{\circ}$. 
It is interesting to note that up to $\theta_{c.m.} \approx 80^{\circ}$ 
 the JISP16 
provides practically the same predictions as the other forces, giving a good 
description of data at this energy. Only above that angle the deviations from 
the other predictions (and data) start to develop. 
At $E=135$ MeV
the JISP16 predictions follow the data at scattering angles 
below $\theta_{\rm{c.m.}}=70^{\circ}$ and above $\theta_{\rm{c.m.}}=150^{\circ}$,
missing the data 
and other predictions at intermediate angles. 
At three lower energies the 3NF effects are negligible and 
the AV18, the AV18+Urbana IX, and N$^4$LO predictions
agree with one another. At $E=135$ MeV the observed 3NF effects are much
smaller than the difference between the JISP16 and the AV18 or 
the chiral results.

\begin{figure}
\includegraphics[width=.8\textwidth,clip=true]{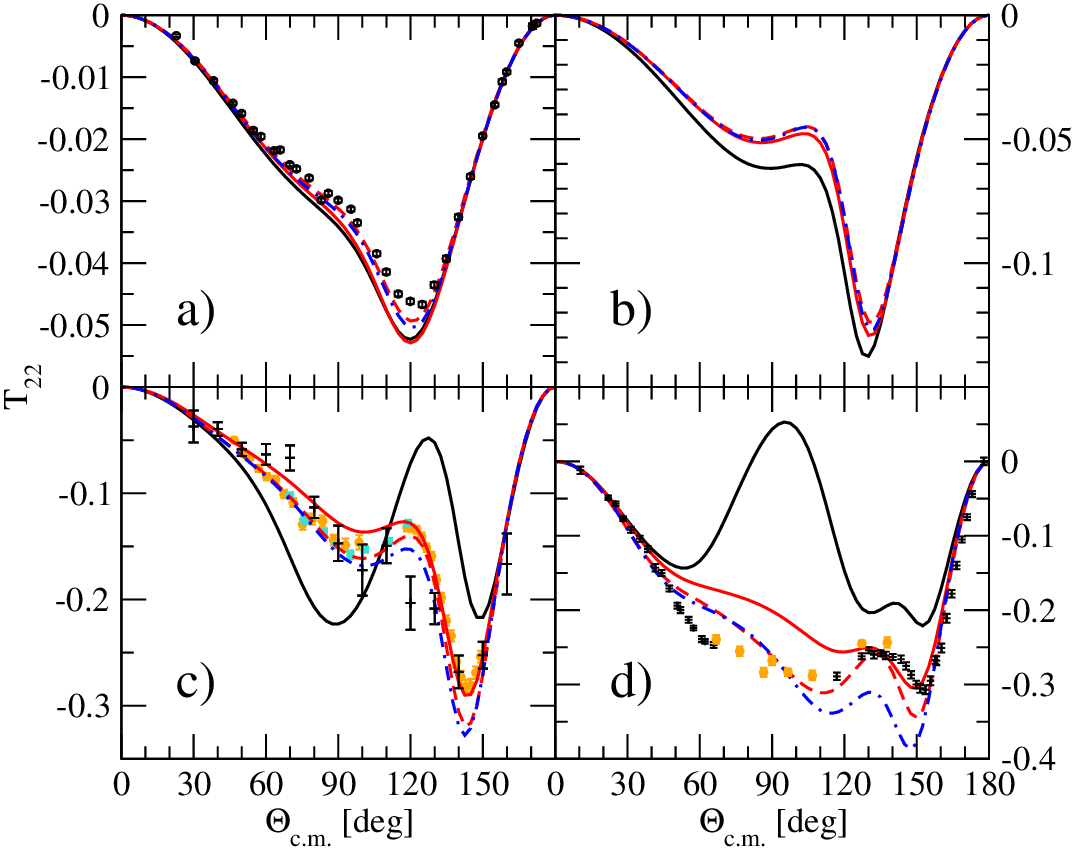}
\caption{(color online) The deuteron tensor analyzing power T$_{22}$ for elastic Nd scattering at the
incoming nucleon laboratory energy a) $E=5$ MeV, b) $E=13 $~MeV, c) $E=65 $~MeV and 
d)~$ E=135 $~MeV as a function of the center-of-mass
scattering angle $\theta_{\rm{c.m.}}$. Curves are as in Fig.~\ref{fig2}.
The data in a) are from Ref.~\cite{Sowinski} ($pd$ circles), in c) are from Ref.~\cite{Witala_it11} ($pd$ pluses), 
~\cite{Stephan} ($pd$ orange circles) and~\cite{Mardanpour} ($pd$ turquoise squares), 
and in d) are from
Ref.~\cite{exp135_1} ($pd$ pluses) and~\cite{exp135_2} ($pd$ orange circles).
}
\label{fig7}
\end{figure}

The tensor analyzing power T$_{22}$, 
presented in Fig.~\ref{fig7}, again shown as a representative of 
observables arising from tensor polarization states, 
belongs to the class of observables that have been proven, at higher energies, to be sensitive 
to fine details of the interaction model.
Predictions based on the JISP16 potential start to deviate slightly 
from the others already at $E=13 $~MeV but
at ${E=65 }$~MeV this deviation becomes large, resulting in a major disagreement 
with the data above $\theta_{c.m.} \approx 30^{\circ}$. 
At $E=135$~MeV  the JISP16 interaction is not able
to describe the data, showing an unexpected maximum 
at $\theta_{\rm{c.m.}}=95^{\circ}$.
At the two higher energies 3NF effects are clearly visible, however 
supplementing the AV18 interaction with the Urbana IX 3NF
moves predictions at $E=135 $~MeV farther away from the data. Even on the level 
of two-body interactions the difference
between the AV18 and the N$^4$LO predictions is large for scattering 
angles around 130$^{\circ}$ and the AV18 
results provide the best description of the data.

Summarizing above results as well as results for remaining, not shown here, elastic scattering observables, 
we can state that the modern nuclear forces, including the JISP16 model, have 
problems with a precise description of
many polarization observables. Moreover,
we conclude that the predictions of the Nd elastic scattering observables
obtained with the JISP16 force are not closer to results arising from the
AV18 plus Urbana IX interactions than to the results obtained when 
 using only NN forces.
In addition, the JISP16 model predictions miss the data  
in broad ranges of the scattering angles. This is seen very clearly 
 at higher energies,
but for some observables also at the relatively small energy of 5 MeV.  

\begin{figure}
\includegraphics[width=.8\textwidth,clip=true]{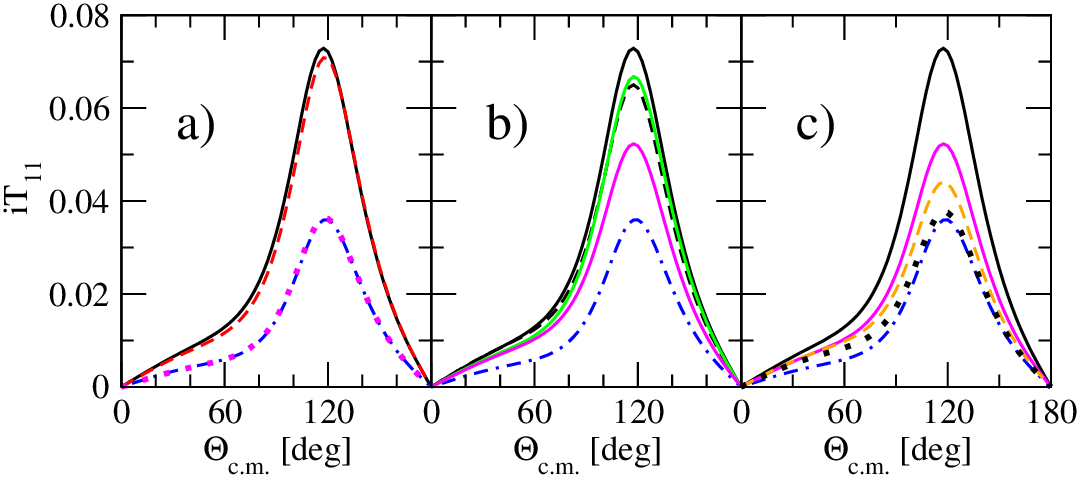}
\caption{(color online) The deuteron analyzing power iT$_{11}$ at the 
incoming nucleon laboratory  energy $E=5$ MeV obtained with different 
inputs to Eq.~(\ref{eq_Fadd_NN}). In all panels the black solid and blue dash-dotted curves are the same as in Fig.~\ref{fig4}
and represent the JISP16 and chiral N$^4$LO predictions, respectively. 
In a) the magenta dotted curve shows results obtained with the chiral N$^4$LO force but using the JISP16 deuteron wave function.
The predictions obtained with the JISP16 force with exception of $^3S_1-{^3D_1}$ $t$-matrix components which
were replaced by the chiral N$^4$LO ones are shown by a red dashed curve.
In panel b) the green solid, black dashed and magenta solid curves represent JISP16 calculations with 
the exchanged $t$-matrix components in the $^1P_1$, $^3P_0$ and $^3P_2-{^3F_2}$ channels, respectively. Finally, in panel c) 
different combinations of the $t$-matrix partial wave components are exchanged: $^3P_2-{^3F_2}$ [the solid magenta curve, the same as in b)],
$^3P_0$ and $^3P_2-{^3F_2}$ (the dashed orange curve) and $^1P_1$, $^3P_0$ and $^3P_2-{^3F_2}$ (the dotted black curve). 
}
\label{fig8}
\end{figure}

Thus the question arises: what is the reason for such a behavior?
We already noticed that the JISP16 deuteron wave function is shifted towards lower momenta compared 
with the wave functions calculated using the other interaction models. 
Since the deuteron wave function is an important ingredient of 
elastic Nd scattering, it is worth investigating if the 
(shifted to small momenta) deuteron wave function can be linked with 
the observed discrepancies. 
To this end we performed calculations with the chiral N$^4$LO semi-locally regularized interaction
(with the regulator $R=0.9 $~fm) but to calculate observables we replaced the chiral N$^4$LO 
deuteron wave function by the one obtained from the JISP16 model.
When calculating the chiral $t$-matrix the correct chiral wave 
function was used to determine the residue of the $^3S_1-^3D_1$ $t$-matrix. 
Having in mind a small deviation of the mixing parameter $\epsilon$ in the $^3S_1-^3D_1$ partial wave
we also repeated calculations with the JISP16, but used the chiral $t$-matrix in that coupled channel instead of the
original JISP16 $t$-matrix.  
Again, the correct residue of the $^3S_1-^3D_1$ $t$-matrix,
given by the chiral deuteron wave function was used. Results at $E=5 $~MeV are shown in panel a) of Fig.~\ref{fig8}.
We focus there on the analyzing power iT$_{11}$, for which the discrepancy between the JISP16 predictions and those of the chiral interaction
is very pronounced. 
As is seen in Fig.~\ref{fig8}a) the iT$_{11}$ practically does not change when the JISP16 
deuteron wave function is used together with the chiral NN force (magenta dotted vs blue dash-dotted curves). 
Also there is only a tiny effect due to changing the $^3S_1-^3D_1$ $t$-matrix (black solid vs red dashed curves). 
Analyzing the importance of the remaining partial waves for this observable we 
found, that the interference of different P-waves is responsible for the observed difference between 
the JISP16 and chiral predictions. This is documented in the middle and in the right panels of Fig.~\ref{fig8}.
Namely, in the middle panel we show what happens if only a single partial wave (channel) component is 
replaced in the $t$-matrix. 
While exchanging most of channels has only a small effect on the iT$_{11}$, 
exchanging separately $^3P_0$, $^1P_1$, or $^3P_2-{^3F_2}$ waves
leads to a significant change in the iT$_{11}$ magnitude (see dashed black, solid green and solid magenta curves respectively).
However, it is clear, that none of these waves is able alone to explain the difference between
the JISP16 and the chiral N$^4$LO predictions. The right panel in Fig.~\ref{fig8} shows effects of 
exchanging two or three partial waves at the same time. We see that the simultaneous exchange of 
the lowest P-waves: $^3P_0$, $^1P_1$ and $^3P_2-{^3F_2}$ shifts predictions close to the 
N$^4$LO results. We conclude, that an interplay 
of these 
partial waves during solving Eq.~(\ref{eq_Fadd_NN}) enhances reduction of the iT$_{11}$ values. 
Of course, the deuteron vector analyzing power at low energies is known to be sensitive to the P-waves but 
results shown in Fig.~\ref{fig8}
clearly indicate that the strength of the P-waves components of the JISP16 interaction should be corrected.   
 
The observation drawn from Fig.~\ref{fig8} is also true for the remaining observables at the low energies:
the replacement of the P-waves components from the JISP16 potential by the 
corresponding ones generated by the N$^4$LO interaction moves the JISP16 based predictions
into the vicinity of the complete chiral N$^4$LO results. 

\begin{figure}
\includegraphics[width=.8\textwidth,clip=true]{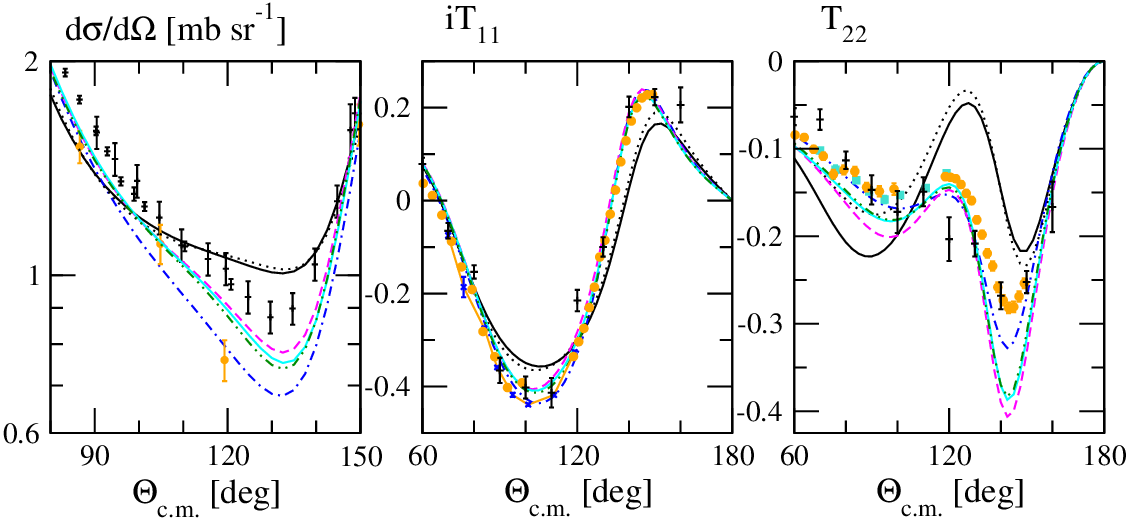}
\caption{(color online) The differential cross section (left panel), the deuteron analyzing power iT$_{11}$ (the middle panel) and 
the deuteron tensor analyzing power T$_{22}$ at the incoming nucleon 
 laboratory  energy $E=65$ MeV obtained with different
inputs to Eq.~(\ref{eq_Fadd_NN}). In all panels the black solid and blue dash-dotted curves are the same as in Figs.~\ref{fig2}-
\ref{fig7}
and represent the JISP16 and chiral N$^4$LO predictions. 
In addition, there are three curves representing 
predictions obtained with 
the JISP16 force but replacing selected sets of the $t$-matrix partial wave components with those
taken from the chiral N$^4$LO potential: 
$^1P_1$, $^3P_0$ and $^3P_2-{^3F_2}$ (black dotted), 
all partial waves with $j \leq 2$ (magenta dashed) and 
all partial waves with $j \leq 3$ (cyan solid).
Finally, the green dash-double-dotted curve shows predictions with the chiral N$^4$LO potential but with
the deuteron wave function generated by the JISP16 NN force. The displayed data follow 
Figs.~\ref{fig2},~\ref{fig4} and~\ref{fig7}
for the cross section, the iT$_{11}$ and the T$_{22}$, respectively.
}
\label{fig8a}
\end{figure}

The picture becomes more complex at the higher energies, where the importance of higher partial waves grows. 
The simultaneous replacement of only the $^1P_1$, $^3P_0$ and $^3P_2-{^3F_2}$ JISP16 $t$-matrix elements by those from the N$^4$LO  
is insufficient to move predictions to the vicinity of the complete chiral results. In Fig.~\ref{fig8a} we show
the differential cross section, the deuteron vector analyzing power iT$_{11}$ and the  deuteron tensor
analyzing power T$_{22}$ at $E=65$~MeV as examples of this different behavior under the change 
of the selected partial wave components of the $t$-matrix.
The range of scattering angles presented in Fig.~\ref{fig8a} is restricted to the regions where the 
differences between various predictions are most noticeable.

The left panel of Fig.~\ref{fig8a} shows the minimum of the differential cross section. The change of
the P-waves (black dotted curve) has only a small influence on the magnitude of the cross section.
A subsequent replacement of all $t$-matrix components up to the two-body total angular momentum $j \leq 2$ (the dashed magenta curve) 
moves predictions in the direction of the N$^4$LO results but this process stops with an increasing
number of replaced partial wave components, as is visible from the cyan solid curve, 
which shows results with all $t$-matrix partial components exchanged up to $j \leq 3$. 
The deuteron wave function and its propagation in the solution of the Faddeev equation can explain most of the 
remaining differences between predictions with the substituted partial wave components and complete N$^4$LO results.
Namely, we show also (by green double-dot-dashed curve) the results obtained with the chiral N$^4$LO 
$t$-matrix combined with the deuteron wave function from the JISP16 force, which are reasonably close 
to the cyan solid curve. 
For the deuteron vector analyzing power, shown in the middle
panel, 
the exchange of only the lowest P-wave components does not affect
the predictions substantially. Only the change of the $t$-matrix in all partial wave components
with $j \leq 2$ moves 
predictions close to the chiral results.
Substituting the higher partial wave components does not change results significantly. The remaining 
differences are due to the different deuteron wave functions and are visible near the minimum 
of the iT$_{11}$. 
In the case of the deuteron tensor analyzing power T$_{22}$ (the right panel) the same modification
of all the partial wave components with $j \leq 2$
does not reproduce the chiral results either. While the maximum of the T$_{22}$ observed for the JISP16 force
around the scattering angle $\theta_{c.m.}=130^{\circ}$ is substantially reduced, the emerging
minimum of the T$_{22}$ at the scattering angles $140^\circ \leq \theta_{c.m.} \leq 150^\circ$ is clearly too deep.
Moreover, even the replacement of the higher partial wave components 
(up to $j \leq 3$) does not remedy this situation. Again the role of the deuteron wave function is important.

\begin{figure}
\includegraphics[width=.8\textwidth,clip=true]{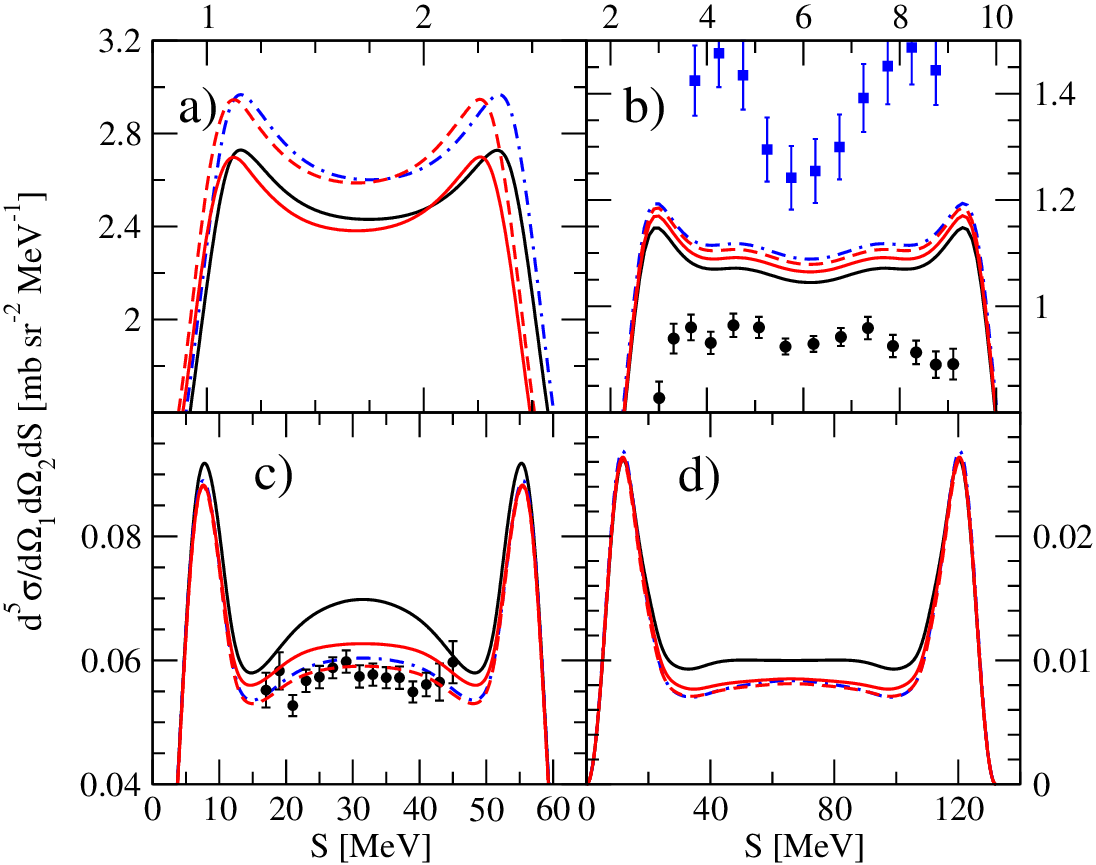}
\caption{(color online) The differential cross section ${\rm d^5}\sigma / {\rm d}\Omega_1 \Omega_2 {\rm dS}$ 
for the SST configuration in the deuteron breakup process at the
incoming nucleon laboratory energy a) E=5 MeV, b) E=13 MeV, c) E=65 MeV
and d) E=135 MeV as a function of the arc length $S$ of the S-curve~\cite{Glockle_raport}.
The polar angles of momenta of two measured neutrons are 
$\Theta_1 = \Theta_2=39.2^{\circ}$ at E=5 MeV,
$\Theta_1 = \Theta_2=50.5^{\circ}$ at E=13 MeV,
$\Theta_1 = \Theta_2=54.0^{\circ}$ at E=65 MeV and
$\Theta_1 = \Theta_2=54.4^{\circ}$ at E=135 MeV,
while the relative azimuthal angle 
$\Phi_{1-2}=120^{\circ}$ at all energies.
Curves are as in Fig.~\ref{fig2}. The data in b) are from Refs.~\cite{Strate88,Strate89} ($nd$ blue squares) 
and from Ref.~\cite{Rauprich} ($pd$ black circles), and in c) are from Ref.~\cite{Zejma} ($pd$ circles).
}
\label{figbreakupSST}
\end{figure}

Finally, we would like to comment briefly on the behaviour of the JISP16 force in the 
deuteron-breakup reaction. In general, the picture obtained with the JISP16 interaction for the deuteron breakup 
resembles the one for the elastic scattering. 
Beside the kinematical configurations in which predictions 
for the exclusive cross section based on
the JISP16 force are in quantitative agreement with predictions of other interactions even at higher energies,
there are configurations in which the JISP16's results clearly differ from remaining predictions.
What is interesting, one of such configurations is the SST configuration~\cite{Glockle_raport}, 
for which cross section is known to be only slightly sensitive to the choice of the NN interaction.
In Fig.\ref{figbreakupSST} we show the exclusive SST differential cross section 
${\rm d^5}\sigma / {\rm d}\Omega_1 \Omega_2 {\rm dS}$ 
at 
energies 
used above for the elastic scattering.
The JISP16 predictions differ from those obtained with 
the other NN interactions: at E=5 and E=13 MeV the differential cross section based on the JISP16 force is 
below other predictions obtained with two-body forces 
(approx. by 7\% and 4\% in the centre of plateau at E=5 MeV and E=13 MeV, respectively) 
while at higher energies it exceeds other results
(approx. by 15\% and 18\% in the centre of plateau at E=65 MeV and E=135 MeV, respectively). 
At E=5 MeV the JISP16 predictions are close to ones for the AV18+UrbanaIX model what can reflect the 
incorporation of 3NF effects into the JISP16 interaction. Also at higher energies results 
with the JISP16 are shifted, compared to the other NN predictions, in the same direction as 
the AV18~+~Urbana~IX results.
However, at the two higher energies this shift is too strong.
Such a picture of the JISP16 SST cross section behaviour
is one more hint that this force
requires a careful revision as already have been concluded from our analysis of the Nd elastic scattering.

We can summarize our findings to this stage by 
a) pointing to the necessity of improvement of the P-wave components in the JISP16 NN potential model 
and 
b) asking a more general question about usefulness of soft potentials in a description of nuclear reactions at intermediate energies.
In the next section we explore the latter issue by studying Nd scattering with a $V_{{\rm low}\;{\rm k}}$ potential.

\section{\boldmath$\rm Nd$ scattering --- JISP16 and low momentum potentials}
\label{Vlowk}

In the following we check whether softening the potential, of course within 
reasonable limits,
could lead to problems with a description of Nd 
scattering 
observables.
If it is possible to construct a NN potential whose matrix elements 
in momentum space 
can be restricted to low momenta and which at the same time guarantees 
a good description 
of observables in few-nucleon reactions, such a force would have very 
welcome properties 
from the point of view of nuclear structure calculations
and lead to reduced computational costs.
Thus in this section we compare predictions obtained 
with the CD Bonn and the JISP16 
NN potentials to
ones based on the $V_{\rm{low}\;{\rm k}}$ potential derived from the CD Bonn force.
We use the cutoff values $\Lambda=1.0$ 
(for the deuteron bound state only), 1.5, 2.0 and 5.0~fm$^{-1}$.

\begin{figure}
\includegraphics[width=.4\textwidth,clip=true]{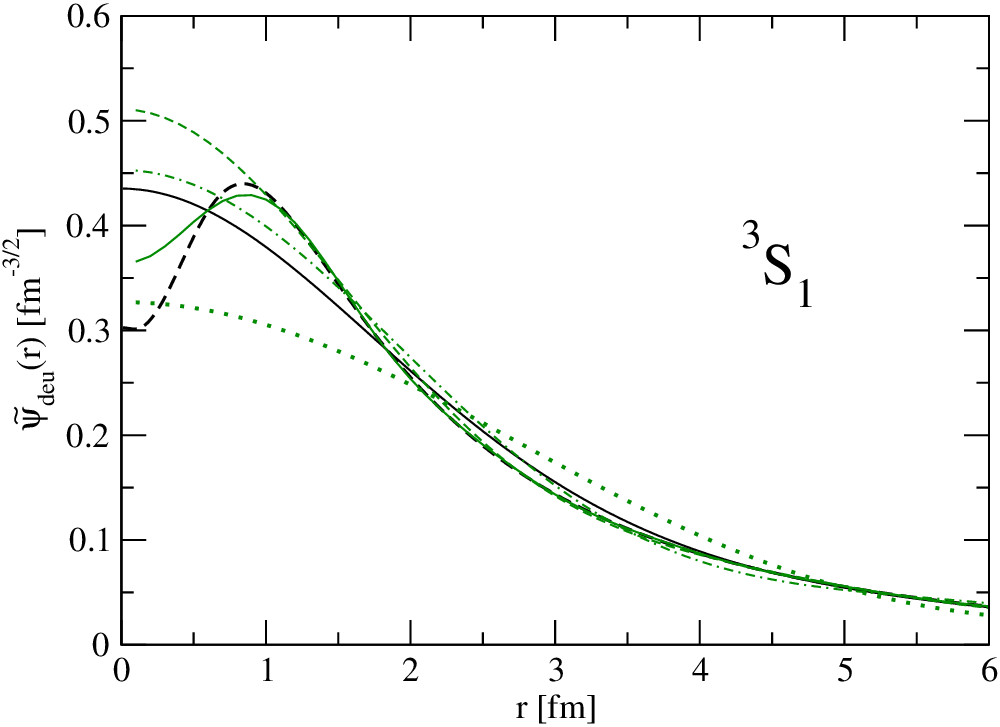}
\includegraphics[width=.4\textwidth,clip=true]{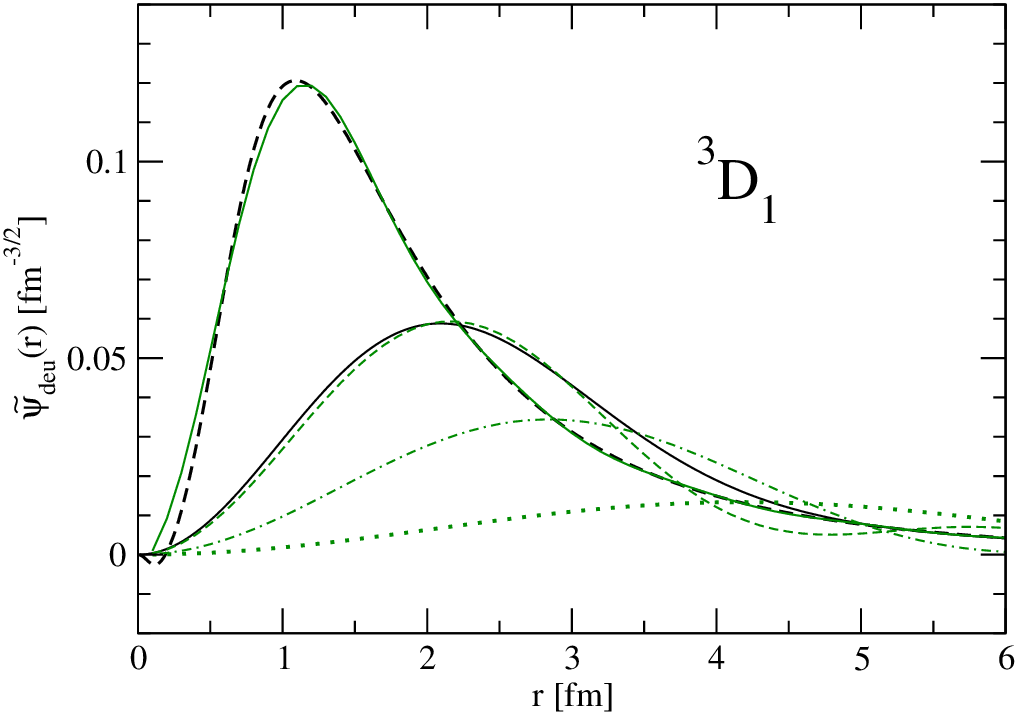}
\includegraphics[width=.4\textwidth,clip=true]{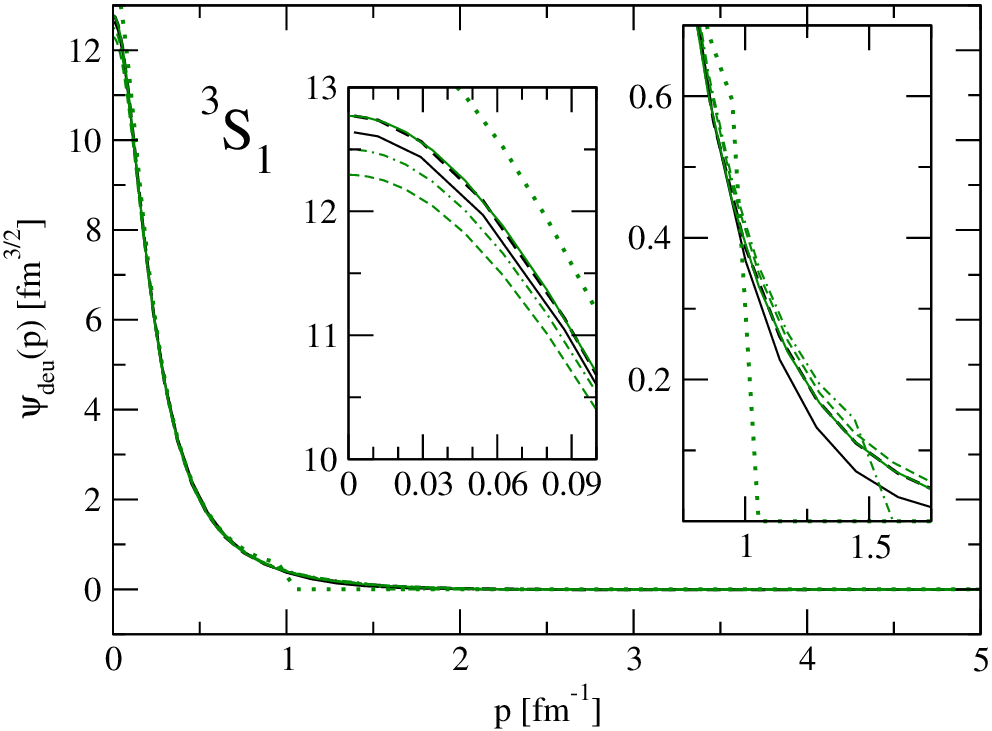}
\includegraphics[width=.4\textwidth,clip=true]{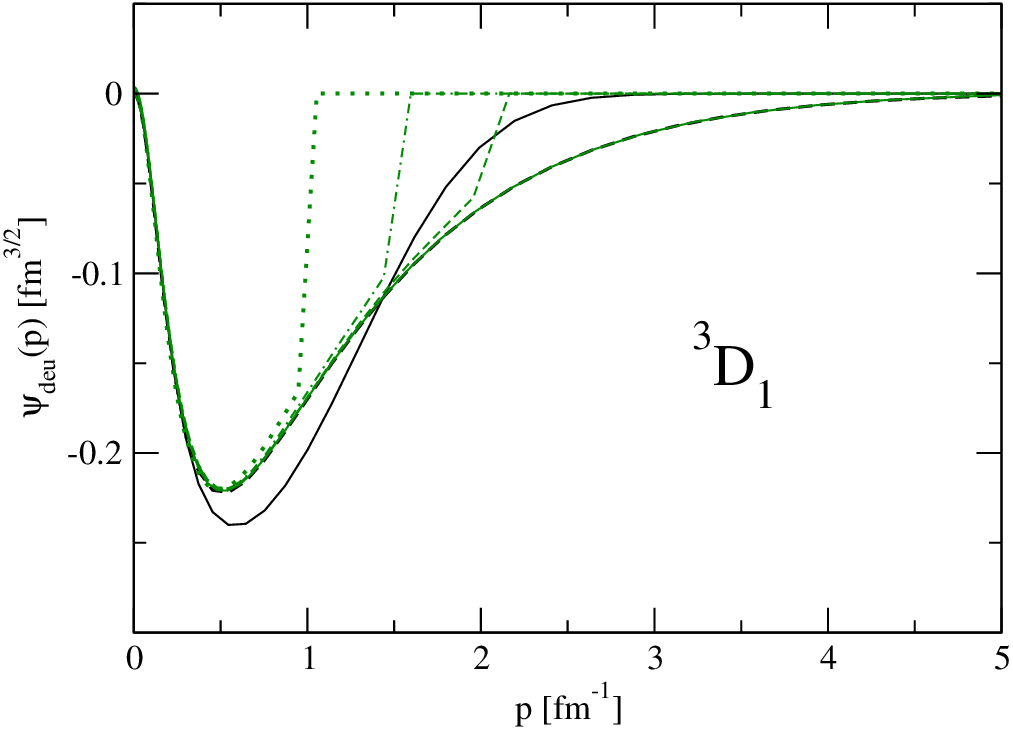}

\caption{(color online) The deuteron wave functions $\tilde{\psi}_{deu}(r)$ (top) and $\psi_{deu}(p)$ (bottom) in
coordinate and momentum space, respectively. The $^3S_1$ and the $^3D_1$ components are shown
in the left and the right column, respectively. The black solid and the black dashed 
curves 
represent the JISP16 and the CD Bonn predictions. The green dotted, dash-dotted, dashed and solid curves show wave functions 
obtained with the $V_{\rm{low}\;{\rm k}}$ interaction based on the CD Bonn force, with cutoff values 
$\Lambda=$ 1.0, 1.5, 2.0 and 5.0~fm$^{-1}$, respectively.}
\label{fig9}
\end{figure}

The corresponding deuteron wave functions are shown in Fig.~\ref{fig9}. 
 At the largest value of $\Lambda$=5~fm$^{-1}$ the two components, $^3S_1$ and $^3D_1$, 
 of the deuteron wave function obtained with this  $V_{\rm{low}\;{\rm k}}$ force 
are very close to the original CD Bonn results. They both have a characteristic 
maximum in coordinate space at $r=1$~fm. 
Changing $\Lambda$ to smaller values, the $^3S_1$ wave function in 
coordinate space loses that maximum and 
monotonically decreases with 
$r$,  similarly to the JISP16 coordinate space $^3S_1$ deuteron component.
In 
momentum space, the same convergence pattern with $\Lambda$ is seen and 
in addition a clear limitation of nonzero wave function components to  
momenta below  $\Lambda$ is observed. 
The probability values
for both components of the deuteron wave 
function
together with the expectation
values of kinetic and potential energies are 
given in Tab.~\ref{tab3}. 

The JISP16 interaction is designed as a matrix in the oscillator basis covering
the NN relative motion up to momenta of approximately 2~fm$^{-1}$. Therefore it is not surprising 
that the JISP16 NN interaction has much in common with the $V_{\rm{low}\;{\rm k}}$ forces, and indeed the
 JISP16 deuteron wave function is seen in Fig.~\ref{fig9} to be close to that from 
 the $V_{\rm{low}\;{\rm k}}$ 
potential with $\Lambda=2\rm~fm^{-1}$. Note however that 
 the kinetic and potential 
energy expectation values in the deuteron given in Tab.~\ref{tab3} indicate that some features
of the JISP16 are closer to those of the $V_{\rm{low}\;{\rm k}}$  with $\Lambda=1.5\rm~fm^{-1}$. 

\begin{table}[b!]
\begin{tabular}{|c|c|c|c|c|c|}
\hline
 & $E_{\rm deu}$~[MeV] & P($^3S_1$) & P($^3D_1$) & $\langle E_{\rm pot} \rangle$~[MeV] & $\langle E_{\rm kin}\rangle$~[MeV] \\
\hline
JISP16                         & $-2.225 $ & 96.02  & 3.98  & $-12.99 $ & 10.76 \\
CD~Bonn (non-rel)             & $-2.223 $ & 95.14  & 4.86  &$ -17.82$  & 15.60 \\
$V_{\rm{low}\;{\rm k}}$ with $\Lambda$=5.0 fm$^{-1}$  & $-2.223 $ & 95.17  & 4.83  &$ -17.53   $& 15.31 \\
$V_{\rm{low}\;{\rm k}}$ with $\Lambda$=2.0 fm$^{-1}$  & $-2.223  $& 96.65  & 3.55  & $-14.22 $  & 12.00 \\
$V_{\rm{low}\;{\rm k}}$ with $\Lambda$=1.5 fm$^{-1}$  & $-2.223  $& 97.48  & 2.52  & $-12.90  $ & 10.68 \\
$V_{\rm{low}\;{\rm k}}$ with $\Lambda$=1.0 fm$^{-1}$  & $-2.223$  & 98.79  & 1.21  & $-11.13$   & 8.81  \\
\hline
\end{tabular}
\caption{The deuteron g.\:s.  
energy $E_{\rm deu}$, the $^3S_1$ and $^3D_1$ state probabilities as well as the potential and the kinetic 
energy expectation values obtained with 
 different NN potentials: JISP16, CD Bonn and $V_{\rm{low}\;{\rm k}}$ derived 
from the CD~Bonn force.}
\label{tab3}
\end{table}

\begin{table}
\begin{tabular}{|c|c|c|c|}
\hline
   & $E_{\rm 3H}$~[MeV] & $\langle E_{\rm pot}^{(NN)} \rangle$~[MeV]  & $\langle E_{\rm kin} \rangle$~[MeV]  \\
\hline
JISP16             &$ -8.37$  &$ -35.77$ & 27.40 \\
$V_{\rm{low}\;{\rm k}}$ with $\Lambda$=1.5 fm$^{-1}$ & $-8.97$ &$ -37.82 $& 28.85   \\
$V_{\rm{low}\;{\rm k}}$ with $\Lambda$=2.0 fm$^{-1}$ &$ -8.84$ & $-40.31 $& 31.47  \\
$V_{\rm{low}\;{\rm k}}$ with $\Lambda$=5.0 fm$^{-1}$ & $-8.27$ & $-45.59$ & 37.32   \\
CD~Bonn                                              & $-8.25$ & $-46.40$ & 38.15 \\
\hline
\end{tabular}
\caption{The $^3$H binding energy $E_{\rm{3H}}$ and the expectation values 
for the 2N potential energy ($E_{\rm pot}^{(NN)}$) and the kinetic 
energy ($E_{\rm kin}$), obtained with various NN interactions only (taken as a neutron-proton force).} 
\label{tab4}
\end{table}

Before we present our 3N scattering results,
 let us emphasize  that the procedure 
that we apply 
to soften the CD Bonn potential and to construct the low momenta  
$V_{\rm{low}\;{\rm k}}$ NN forces (i.e by applying unitary 
transformations with different 
values of the cutoff parameter $\Lambda$)  preserves the 
 good description of 
the NN system 
given  by the CD Bonn potential.
This is exemplified by the equal values of the 
deuteron g.\:s. 
energy $E_{\rm deu}$ given in Tab.~\ref{tab3} for the CD~Bonn 
and different versions of $V_{\rm{low}\;{\rm k}}$. 
 However, that 
procedure, when applied to a 3N system generates additional   
3N forces \cite{jurgenson2009,furnstahl2013}, which, generally speaking,
 should  be included when performing 3N calculations. 
Neglecting those additional 3NF's can lead to misleading conclusions,  
with the exception of cases when the 3NF effects are negligible. 
From an approximate equality of the triton binding energies of the CD~Bonn and 
 $V_{\rm{low}\;{\rm k}}$ NN force with  $\Lambda$=5~fm$^{-1}$ (see Tab.~\ref{tab4}) 
it follows, that the effects of missing 3N forces are practically negligible for  that 
observable at $\Lambda$=5~fm$^{-1}$. However, decreasing 
the cutoff $\Lambda$ to 
$\Lambda$=2~fm$^{-1}$ and $\Lambda$=1.5~fm$^{-1}$, makes 
 the additional 3N interaction 
indispensable, since the $V_{\rm{low}\;{\rm k}}$ predictions for the triton binding energy
are significantly higher (by $\approx 0.7$~MeV) than the CD~Bonn results. 
 Including the induced 
3N force into the triton calculations should regain
the CD~Bonn result as was shown in  Ref.~\cite{jurgenson2009} for the case of 
the chiral N$^3$LO potential and a number of low momenta forces 
generated from that interaction.
 
\begin{figure}[b!]
\includegraphics[width=.78\textwidth,clip=true]{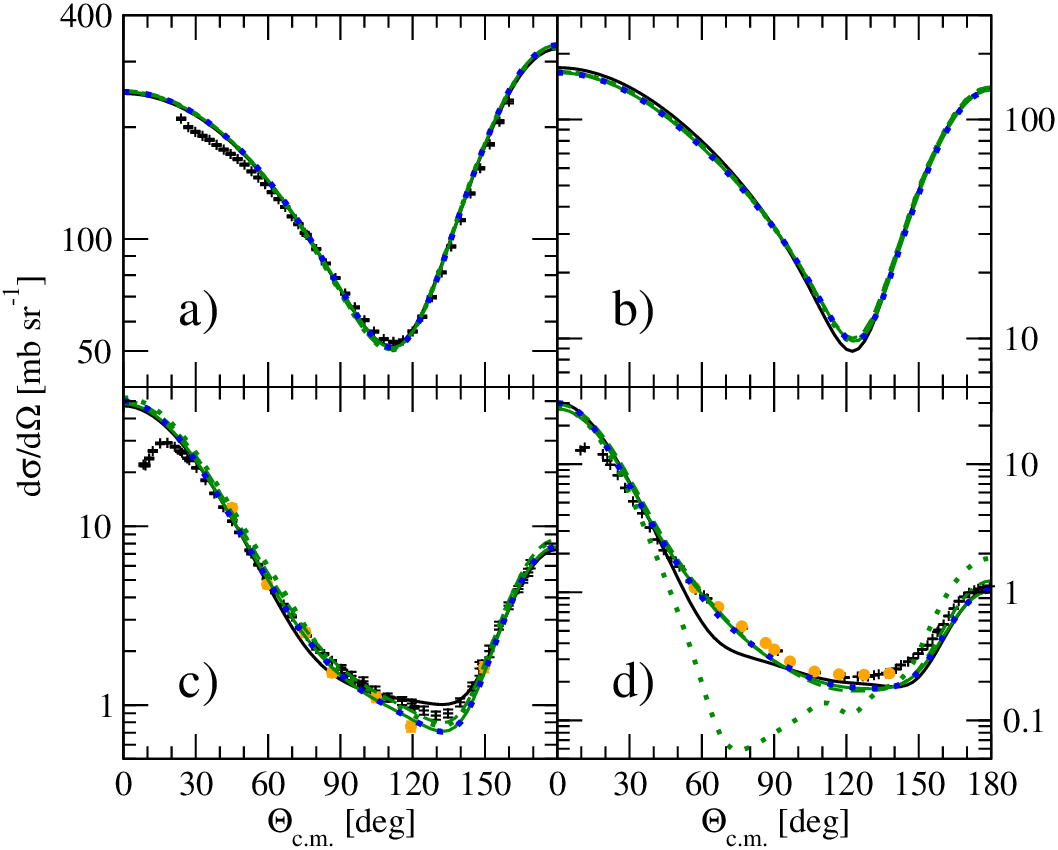}
\caption{(color online) The same as in Fig.~\ref{fig2} but for the theoretical predictions based on the $V_{\rm{low}\;{\rm k}}$ 
interaction with the
cutoff parameters $\Lambda=5$~fm$^{-1}$ (the green solid curve), $\Lambda=2\rm~fm^{-1}$ (the green dashed curve)
and $\Lambda=1.5\rm~fm^{-1}$ (the green dotted curve). The black solid curve represents the JISP16 
predictions and is the same as in Fig.~\ref{fig2}. The blue dotted curve is for the CD Bonn based results
and overlaps with the green solid one.}
\label{fig10}\vspace{-3ex}
\end{figure}

\begin{figure}
\includegraphics[width=.8\textwidth,clip=true]{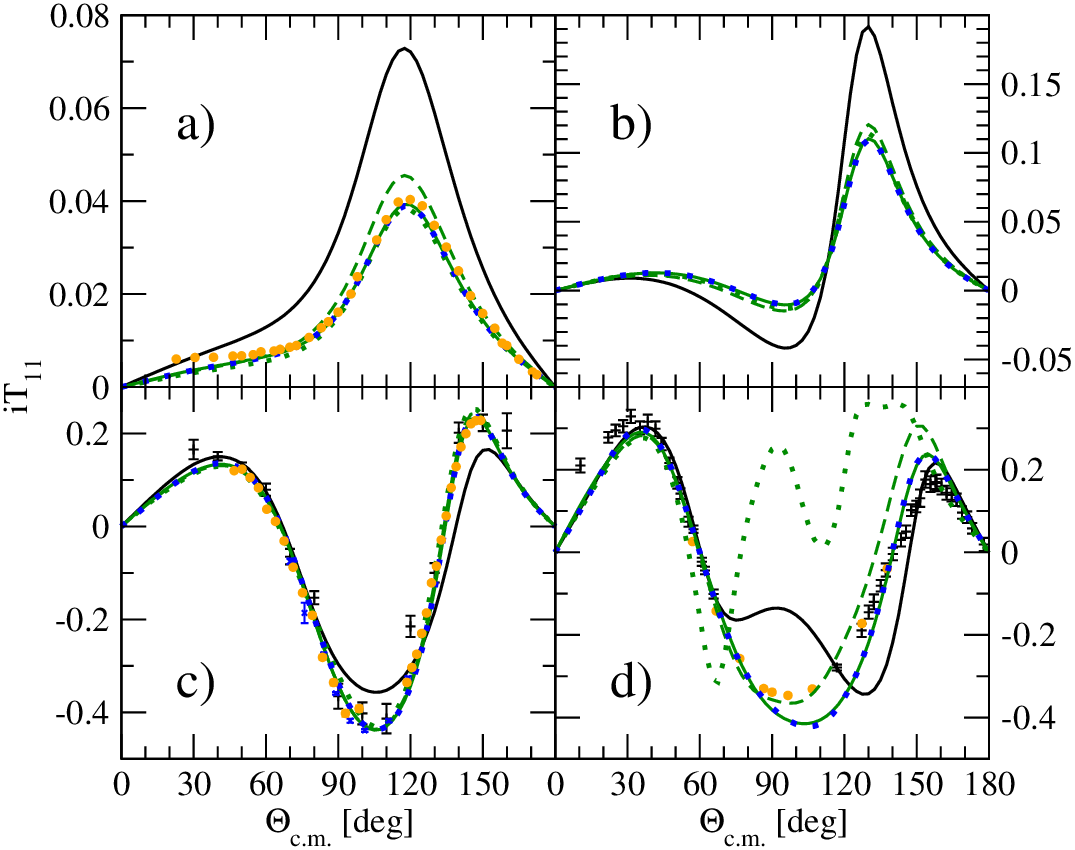}
\caption{(color online) The same as in Fig.~\ref{fig4} but for the theoretical predictions based on the $V_{\rm{low}\;{\rm k}}$ interaction. 
Curves are as in Fig.~\ref{fig10} and data are as in Fig.~\ref{fig4}.} 
\label{fig12}
\end{figure}

\begin{figure}
\includegraphics[width=.8\textwidth,clip=true]{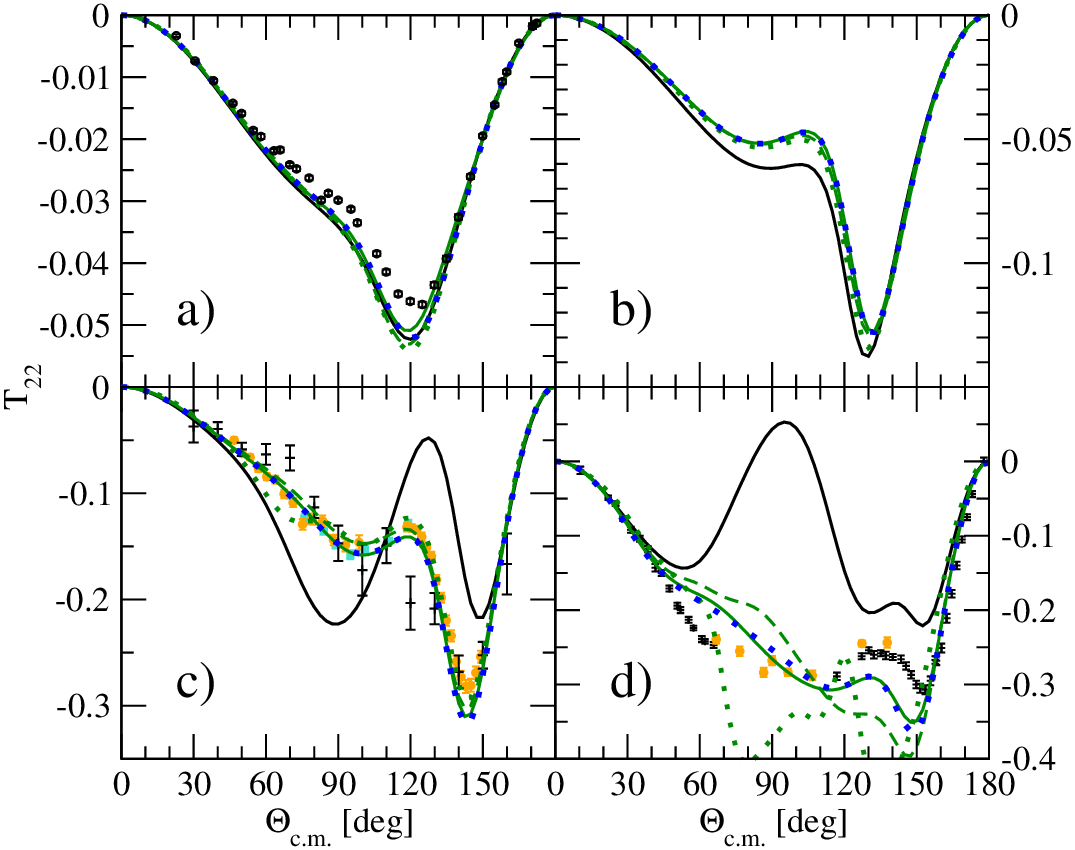}
\caption{(color online) The same as in Fig.~\ref{fig7} but for the theoretical predictions based on the $V_{\rm{low}\;{\rm k}}$ interaction. 
Curves are as in Fig.~\ref{fig10} and data are as in Fig.~\ref{fig7}.} 
\label{fig15}
\end{figure}

For the 3N scattering observables we give again a set of figures 
(Figs.~\ref{fig10}--\ref{fig15})
showing the differential cross section and the vector and tensor 
analyzing powers at the nucleon laboratory energies 
$E=5$, 13, 65 and 135 MeV. Apart from the predictions obtained with 
the $V_{\rm{low}\;{\rm k}}$ potential derived 
from the CD~Bonn NN interaction for various values of the cutoff 
parameter $\Lambda$, we show the reference 
results based on the CD Bonn force, which corresponds to 
the limit $\Lambda \rightarrow \infty$. For the convenience of the 
reader we present again also the JISP16 predictions.

 At the three lower energies $5, 13$, and $65$~MeV 
the differential cross section, shown in Fig.~\ref{fig10}, is 
 stable with respect to changing the $\Lambda$ value and cross sections 
for different cutoffs are close to the CD~Bonn prediction.
 At $E=135$ MeV only the predictions for $\Lambda$=1.5~fm$^{-1}$ deviate 
drastically from the other cutoffs, which are close to the 
 CD~Bonn cross sections. 
 The  $\Lambda=1.5\rm~fm^{-1}$ cross sections show strong variations with the
scattering angle, which cannot be attributed to effects due to action of some 
 3NF, since such effects generally change smoothly with the scattering angle. 
It indicates that as far as
the cross section is
concerned the effects of the additional 3N force are at most moderate.
Further we argue that the observed behavior of the cross
section at 135 MeV and $\Lambda=1.5$~fm$^{-1}$ is dominated by kinematical
restrictions coming
into effect for low momenta interactions.

For polarization observables the picture is similar to the one for 
the cross section although 
they become somewhat more 
sensitive 
to variations of the cutoff parameter at some energies and angles 
(see Figs.~\ref{fig12} and \ref{fig15} 
for representatives of the analyzing powers: $iT_{11}$ and $T_{22}$ ). 
At the two lowest energies, $E=5$~MeV and $E=13$~MeV, predictions for 
the $iT_{11}$ and $T_{22}$ as well as for remaining, not shown here,  
polarization observables  
are practically insensitive to changes of the cutoff parameter. 
At these two energies results for all cutoffs are very close to each 
other and to the CD~Bonn potential prediction in the whole angular region. 
The same is true at $E=65$~MeV for 
the $iT_{11}$ for all $\Lambda$ values. 
At that energy the tensor analyzing power
$T_{22}$, 
 shows only weak dependence on 
the cutoff value 
$\Lambda$
(see Figs.~\ref{fig12} and \ref{fig15} ). 
Overall, this indicates that,
for polarization observables, the effects of 
the additional 3N forces are negligible, at least at the two largest $\Lambda$
values. 

Going to $E=135$~MeV one finds again a picture similar to the one observed for the cross 
section at that energy. Again for 
both polarization observables the 
prediction with the smallest cutoff $\Lambda=1.5$~fm$^{-1}$ becomes 
drastically different from the others for angles above some specific value, 
showing rapid angular variations.
At this energy even predictions for $\Lambda=2$~fm$^{-1}$ start to 
 reveal  such a
behavior which, like the behavior of the cross section at 
that energy, arises mostly from kinematical restrictions.

It is worth emphasizing that for all observables predictions 
based on the $\Lambda=5\rm~fm^{-1}$ are at all four energies 
indistinguishable from the original 
CD Bonn results.

To explain the reason for such a behavior of the low momenta potential 
predictions let us stress
that when using such interactions 
in 3N continuum calculations, 
 a natural limitation 
appears
and results obtained with such forces 
can be applied to interpret elastic Nd scattering data
only up to a certain 
initial relative nucleon-deuteron momentum.
 That limitation follows from the fact that the momentum space 
deuteron wave function components as well as the momentum space 
 matrix elements of such a low momentum  $V_{\rm{low}\;{\rm k}}$ potential are 
restricted to momentum values 
below the cutoff parameter $\Lambda$ of that force.  
It means that application of such forces to interpret 
elastic Nd scattering data  at specific  incoming nucleon laboratory 
energy is possible only in a limited  region of c.\:m. angles where
the momentum transfer $\Delta q=2q_0 \sin \frac {\theta_{c.m.}} {2}$ is  
smaller that $\Lambda$. (Here $q_0$ is the magnitude of the relative 
nucleon-deuteron momentum.) It restricts the application of low momentum potentials 
to the c.\:m. angles below $\theta^{lim}_{c.m.}=2 \arcsin( \frac {\Lambda} {2q_0})$ for 
a given incoming nucleon energy and thus also restricts the incoming nucleon 
laboratory energy to a region below $E_{lab}^{lim}=\frac {9} {32m} \Lambda^2$
($m$ is the nucleon mass), where predictions should be valid over the whole angular range.  
In Tab.~\ref{tab5} we show that limiting angle at each 
energy studied in the present paper for two 
values of $\Lambda=1.5$~fm$^{-1}$ and 
$\Lambda=2$~fm$^{-1}$. 
 Assuming that 
effects of additional 3N forces are negligible, as we 
inferred for certain kinematic regions from the 
above analysis and what in addition is supported by results 
of Ref.~\cite{Deltuva}, it follows from the numbers given in Tab.~\ref{tab5}, that 
 the low momentum interactions $V_{\rm{low}\;{\rm k}}$ should 
provide an equally good description of Nd elastic scattering 
data as the CD~Bonn potential at the two lowest energies $E=5 $~MeV and ${E=13 }$~MeV. 
 The deviations from the CD~Bonn potential based predictions can appear at 
 $E=65 $~MeV and $E=135 $~MeV, where 
the limiting angle restricts description
 to angles below $\theta_{c.m.}^{lim}$. At these energies, 
especially for $E=135 $~MeV and $\Lambda=1.5$~fm$^{-1}$, where the 
limiting angle is
smallest,  for some observables low momentum interaction predictions
follow the CD~Bonn ones only for angles up to $\approx \theta_{c.m.}^{lim}$, 
diverging
strongly at the larger angles. This is just what is seen in 
Figs.~\ref{fig10}--\ref{fig15}. 
Since the JISP16 behaves in some respects 
similarly to the low momentum $V_{\rm{low}\;{\rm k}}$ potential 
for $\Lambda=2$~fm$^{-1}$,
the above arguments explain also the behavior of its predictions seen in 
 Figs.~\ref{fig10}--\ref{fig15}. 
 One could argue that in order to get a proper description of data by such 
a low momentum interaction in the whole angular range, the following 
 restriction on 
the momentum transfer should be imposed: $2q_0 \le \Lambda$.
This in turn can be used to establish the maximal energy, 
below which low 
momentum potentials can be applied to interpret the full angular range of the Nd  
elastic scattering data, $E_{lab}^{lim}$.
 That restriction  gives for $\Lambda=2$~fm$^{-1}$ the 
 limiting energy of
$E_{lab}^{lim}=46.7$~MeV and the  limiting energy of $E_{lab}^{lim}=26.2$~MeV
for $\Lambda=1.5$~fm$^{-1}$. 
These conclusions are further supported by the (not shown here) results obtained
with the cutoff value $\Lambda=3.5$~fm$^{-1}$ at $E=135$~MeV (at this energy $2q_0=2*1.70$~fm$^{-1}$=3.4~fm$^{-1}$) 
for which we observe a good agreement
between $V_{\rm{low}\;{\rm k}}$ and CD Bonn predictions over the full angular range for all observables.
Thus in any future attempts to improve the JISP16 potential by adjusting its 
parameters to 
the elastic scattering Nd data in addition to NN and many-body nuclear 
structure observables,
one could 
benefit only
from the use of  
low energy Nd data below $E_{lab} \approx 30$~MeV.
In the case of the low-momentum interactions which, unlike the JISP16, require
the additional 3NF the final conclusions on their usefulness to study Nd scattering at higher energies
could be drawn only after performing calculations taking the full Hamiltonian into account.
However, the external scale of momenta given by the momentum of the 
incoming nucleon, which defines the relative momentum scale, imposes
the limit on the energy at which the Nd elastic scattering can be investigated with the $V_{\rm{low}\;{\rm k}}$ forces
at a given cutoff $\Lambda$.

\begin{table}[b!]
\begin{tabular}{|c|c|c|c|c|}
\hline
 ~$E_{lab}$~[MeV]~ & $~q_0$~[fm$^{-1}]~$ & ~$\theta^{lim}_{c.m.}(\Lambda=1.5$~fm$^{-1})$~[deg]~ &  ~$\theta^{lim}_{c.m.}(\Lambda=2~$fm$^{-1})$~[deg]~ \\
\hline
5             & 0.33  & 180.0  &  180.0    \\
13            & 0.53  & 180.0  &  180.0    \\
65            & 1.18  & 78.9 &  115.8    \\
135           & 1.70  & 52.3 &  72.0    \\
\hline
\end{tabular}
\caption{The limiting c.m. angle 
$\theta^{lim}_{c.m.}=2 \arcsin( \frac {\Lambda} {2q_0})$, at 
which the momentum transfer in elastic Nd scattering at the incoming nucleon 
laboratory energy $E_{lab}=\frac {9} {8m} q_0^2$ 
becomes greater than the specified $\Lambda$ value.
The corresponding nucleon relative momentum is denoted as $q_0$ and 
the nucleon mass as $m$.}
\label{tab5}
\end{table}

\begin{figure}
\includegraphics[width=.8\textwidth,clip=true]{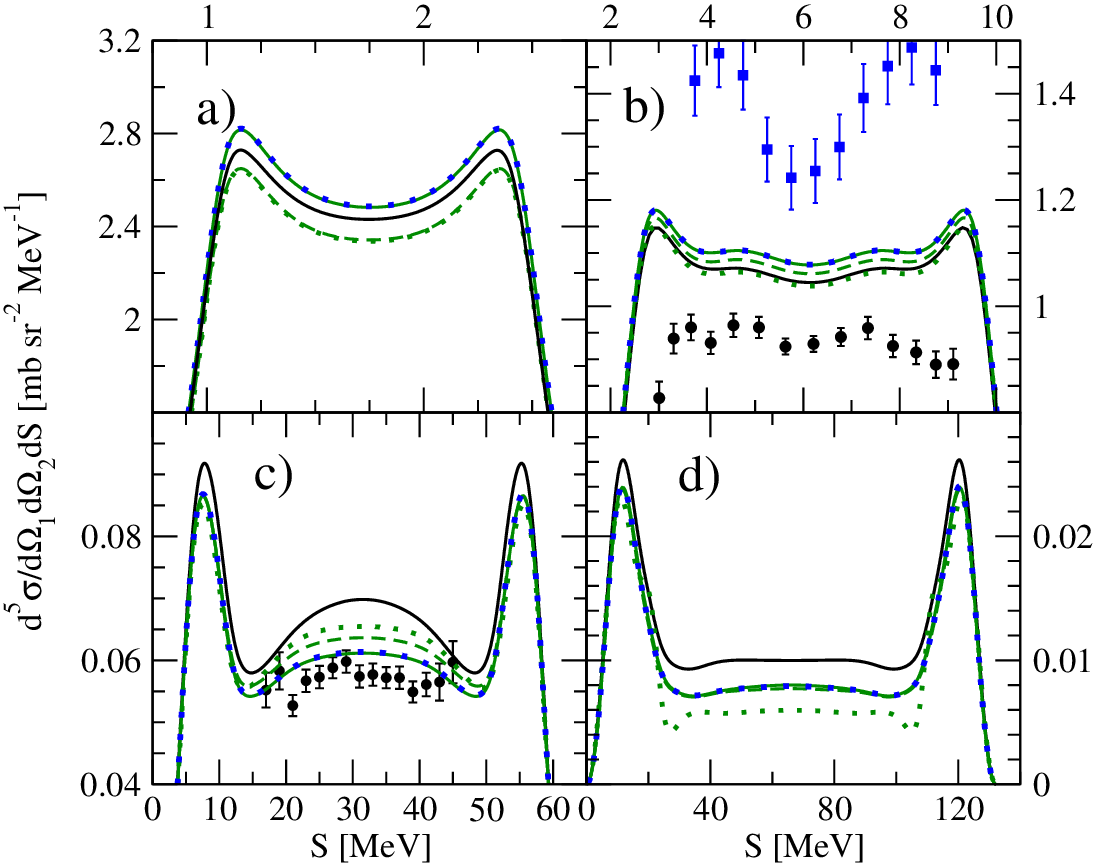}
\caption{(color online) The same as in Fig.~\ref{figbreakupSST} but for 
predictions based on the $V_{\rm{low}\;{\rm k}}$ interaction. 
Curves are as in Fig.~\ref{fig10} and data are as in Fig.~\ref{figbreakupSST}.
} 
\label{figbreakupSSTvlowk}
\end{figure}

Finally, in Fig.~\ref{figbreakupSSTvlowk} we show predictions
for the differential cross section of the SST configuration in the deuteron
breakup process. 
We observe that for all energies predictions obtained with the two lowest values of 
the cutoff $\Lambda$ differ from ones obtained with $\Lambda$=5 fm$^{-1}$,
which are practically indistinguishable from the CD-Bonn results.
The JISP16 predictions at the two lowest energies are close to those arising from
the $V_{\rm{low}\;{\rm k}}$ with the cutoff value $\Lambda$ between 1.5 and 3.0 fm$^{-1}$.
At E=65 MeV and E=135 MeV the JISP16 model gives
cross sections significantly above other results.  
We can conclude that, as for elastic scattering, 
also in the deuteron breakup we observe qualitatively similar behaviour 
of predictions based on the JISP16 or $V_{\rm{low}\;{\rm k}}$ models. 

\section{Summary and conclusions}
\label{Summary}

In the present work the Nd 
scattering 
process is investigated 
for the first time with
the JISP16 NN interaction, which has already been successfully applied
to nuclear structure studies. This soft interaction was constructed with 
the aim of incorporating genuine 3NF effects 
via phase-shift equivalent modifications of the 2N potential, thereby 
simplifying extremely complex many-body calculations of nuclear structure
by reducing the need for explicit 3NFs.

The 
comparison of the deuteron properties
for the JISP16 force and the low momentum interactions $V_{\rm{low}\;{\rm k}}$ 
obtained from
the CD Bonn force within the framework of unitary 
transformation~\cite{Fujii}, reveals the similarity of both models
when the cutoff parameter $\Lambda \approx$2~fm$^{-1}$ is used for 
the $V_{\rm{low}\;{\rm k}}$ force.
Thus we also studied Nd elastic scattering and the nucleon induced 
deuteron breakup process with the $V_{\rm{low}\;{\rm k}}$ 
forces for various 
values of the cutoff parameter. 
Decreasing the cutoff parameter $\Lambda$ leads to a growing importance 
 of the induced 
3N force, obtained within the  renormalization group methods, 
when such low momentum interactions are applied in 3N structure calculations 
 \cite{jurgenson2009,furnstahl2013}. Effects of such induced 3N force are 
significant in $^3$H and $^3$He bound state calculations for small values 
of $\Lambda$ \cite{jurgenson2009}. For the 3N continuum  
the effects of additional 3NF related to the $V_{\rm{low}\;{\rm k}}$ force used here
are rather small  
as shown by our comparison of low momentum interaction predictions 
to the CD~Bonn potential results for 
Nd scattering observables. 
That conclusion is 
further supported by results of calculations presented in Ref.~\cite{Deltuva}, 
where induced 3N forces have been included in 3N continuum calculations. 
  The  condition that the momentum transfers in elastic 
Nd scattering  cannot exceed the limiting momentum $\Lambda$ of the 
low momentum interaction  restricts  the application of low momentum 
 interactions to  low 
energies if the full scattering range is to be investigated.
 We have found that 
only for small 
energies of the incoming nucleon, 
the cutoff values in the range 
1.5~fm$^{-1} \leq  \Lambda < 5$~fm$^{-1}$
can be 
used.
At the energies equal or higher than 65 MeV 
the value $\Lambda =5$~fm$^{-1}$ delivers results equivalent to those based 
on the genuine  CD Bonn potential.

Indeed, our results reveal that the application of the JISP16 force 
to a description of 
Nd elastic scattering should be restricted 
to the low energy domain, below approximately 30~MeV. 
In the case of the deuteron breakup the applicability 
of the JISP16 model depends both on the scattering energy and on the 
final kinematical configuration.
Moreover,  
the P-wave components of
this force require improvements
because they lead to
strong discrepancies with data at low energies for elastic Nd scattering 
observables. This is the case
for the deuteron vector analyzing power, which is sensitive to 
the NN interaction in the P-waves. 
With the current version of the JISP16 force one obtains a reasonable 
description of the data 
only for some observables such as the differential cross section or 
the deuteron tensor analyzing power T$_{21}$.
The description of the 3N scattering data obtained with the JISP16 model is not
as satisfactory as the 
 description of nuclear energy levels achieved with 
this interaction.

Comparing the NN~+~3NF results and predictions based on the two-body forces only,
we cannot conclude that the JISP16 results are closer to the predictions 
based on the 
AV18~+~Urbana~IX potentials than results obtained from the other models of 
the NN interaction. So,  
the conclusion based on nuclear structure 
calculations, that the JISP16 minimizes 
the 
genuine 3NF effects must be updated with an
essential addition that
this is true only for some observables and at a limited range of the momentum transfer.
 It should be 
  emphasized  that in the nuclear 
structure calculations the binding energy 
 is produced  through a 
subtle cancellation of kinetic and potential energies while in the
3N reactions one deals with the S-matrix governed by the full potential energy. 
This basic difference is the reason that in the two domains of negative 
and positive energies the 
 3NF comes into play in  
 different ways. This observation
 is supported 
by studies of 3NF effects in the 3N continuum, where the  
magnitude of the genuine 3NF effects seems to be small at low energies. 
 Only when going to  
larger momentum transfers do the essential
 effects of 3NF's appear.
Our results point to the possible role played by the induced
3NF when the $V_{\rm{low}\;{\rm k}}$ force is used.
It would be very interesting to check in future studies, combining a low momentum NN interaction,
induced 3NF and genuine 3NF, to what extent the description of the Nd scattering observables
is recovered in such a treatment.

The successful performance of the JISP16 model in the structure 
calculations and observed deficiencies 
in the scattering studies exemplify the fact that the continuum 
states deliver additional challenging 
tests of the NN potentials. They can bring out the features of 
the interaction which are of less
importance in nuclear structure calculations.
This in turn leads to a conclusion that developing future models 
of nuclear forces, including those
derived using the inverse scattering methods, 
in addition to the nuclear properties, 
the observables in few-nucleon reactions
should also
be taken into account when fixing potential parameters.
We plan to examine these possibilities in future improvement
of a new high-quality NN force Daejeon16~\cite{Dae16} which reproduces
 observables in light nuclei without the use of 3NFs with a better accuracy than JISP16.

\acknowledgments

This work is a part of the LENPIC project.
It was supported by the Polish National Science Center under Grants No. DEC-2013/10/M/ST2/00420,
DEC-2016/21/D/ST2/01120 and by Grant-in-Aid for Scientific
Research (B) No: 16H04377, Japan Society for the Promotion of Science (JSPS).
The work of AMS was supported by the Russian Science Foundation under
project No.~16-12-10048. This work was also supported in part by the US Department of Energy 
under grant DE-FG02-87ER40371.
The numerical calculations were partially performed on the interactive server at
RCNP, Osaka University, Japan, and on the supercomputer cluster of the JSC,
J\"ulich, Germany.

\end{document}